\RequirePackage{fix-cm}
\documentclass[natbib]{svjour3}                     

\smartqed  
\usepackage{graphicx}
\usepackage{epsf}
\usepackage{amssymb}
\usepackage{amsmath}
\usepackage{placeins}
\usepackage{color}

\journalname{SSRv}
\def\citep{\cite}
\def\lsim{\mathrel{\rlap{\lower 4pt \hbox{\hskip 1pt $\sim$}}\raise 1pt
\hbox {$<$}}} 
\def\gsim{\mathrel{\rlap{\lower 4pt \hbox{\hskip 1pt $\sim$}}\raise 1pt
\hbox {$>$}}}
\def\ltsim{$\lsim$ }
\def\gtsim{$\gsim$ }

\newcommand{\be}{\begin{equation}}
\newcommand{\ee}{\end{equation}}
\newcommand{\beq}{\begin{eqnarray}}
\newcommand{\eeq}{\end{eqnarray}}
\newcommand\subsun[1]{{$_{\normalsize\odot}$}}

\def\e#1{$\times$ $10^{#1}$ }
\def\ee#1{$10^{#1}$ }
\def\mdel#1{$\Delta M_{\rm #1}$ }
\def\mdelp#1{$\Delta M_{\rm #1}$}
\def\mdp#1{$\dot M_{\rm #1}$ }

\def\mm#1{$M_{\rm #1}$ }
\def\mmp#1{$M_{\rm #1}$}
\def\nco{$^{14}$N(e$^-$, $\nu$)$^{14}$C($\alpha$, $\gamma$)$^{18}$O }

\begin{document}
\title{Single Degenerate Models for Type Ia Supernovae}

\subtitle{Progenitor's Evolution and Nucleosythesis Yields}

\titlerunning{Single Degenerate Models}        

\author{Ken'ichi Nomoto, Shing-Chi Leung}
\institute{Ken'ichi Nomoto \at
Kavli Institute for the 
Physics and Mathematics of the Universe (WPI), The University of Tokyo,
Kashiwa, Japan
\email{nomoto@astron.s.u-tokyo.ac.jp} \\ \\
Shing-Chi Leung \at
Kavli Institute for the 
Physics and Mathematics of the Universe (WPI), The University of Tokyo,
Kashiwa, Japan
\email{shingchi.leung@ipmu.jp}
}
\date{Received: 8 March 2018 / Accepted: 28 March 2018}

\maketitle


\begin{abstract}
  We review how the single degenerate models for Type Ia supernovae
  (SNe Ia) works.  In the binary star system of a white dwarf (WD) and
  its non-degenerate companion star, the WD accretes either
  hydrogen-rich matter or helium and undergoes hydrogen and helium
  shell-burning.  We summarize how the stability and non-linear
  behavior of such shell-burning depend on the accretion rate and the
  WD mass and how the WD blows strong wind.  We identify the following
  evolutionary routes for the accreting WD to trigger a thermonuclear
  explosion.  Typically, the accretion rate is quite high in the early
  stage and gradually decreases as a result of mass transfer.  With
  decreasing rate, the WD evolves as follows: (1) At a rapid accretion
  phase, the WD increase its mass by stable H burning and blows a
  strong wind to keep its moderate radius.  The wind is strong enough
  to strip a part of the companion star's envelope to control the
  accretion rate and forms circumstellar matter (CSM).  If the WD
  explodes within CSM, it is observed as an ``SN Ia-CSM''.  (X-rays
  emitted by the WD are absorbed by CSM.)  (2) If the WD continues to
  accrete at a lower rate, the wind stops and an SN Ia is triggered
  under steady-stable H shell-burning, which is observed as a
  super-soft X-ray source: ``SN Ia-SSXS''. (3) If the accretion
  continues at a still lower rate, H shell-burning becomes unstable
  and many flashes recur.  The WD undergoes recurrent nova (RN) whose
  mass ejection is smaller than the accreted matter.  Then the WD
  evolves to an ``SN Ia-RN''.  (4) If the companion is a He star (or a
  He WD), the accretion of He can trigger He and C double detonations
  at the sub-Chandrasekhar mass or the WD grows to the Chandrasekhar
  mass while producing and He-wind: ``SN Ia-He CSM''.  (5) If the accreting WD
  rotates quite rapidly, the WD mass can exceed the Chandrasekhar mass
  of the spherical WD, which delays the trigger of an SN Ia.  After
  angular momentum is lost from the WD, the (super-Chandra) WD
  contracts to become a delayed SN Ia.  The companion star has become
  a He WD and CSM has disappeared: ``SN Ia-He WD''.  We update
  nucleosynthesis yields of the carbon deflagration model W7, delayed
  detonation model WDD2, and the sub-Chandrasekhar mass model to
  provide some constraints on the yields (such as Mn) from the
  comparison with the observations.  We note the important metallicity
  effects on $^{58}$Ni and $^{55}$Mn.
\end{abstract}


\bibliographystyle{ieeetr/unsrtnat}

\section{Introduction}
\label{intro}

The thermonuclear explosion of a carbon+oxygen (C+O) white dwarf has
successfully explained the basic observed features of Type Ia
supernovae (SNe Ia) \cite[e.g.,][]{Nomoto2017,LN2017}. Both the
Chandrasekhar and the sub-Chandrasekhar mass models have been examined
\cite[e.g.,][]{Livio2000}.  However, no clear observational indication
exists as to how the white dwarf mass grows until carbon ignition, i.e.,
whether the white dwarf accretes H/He-rich matter from its binary
companion [single-degenerate (SD) scenario] or whether two C+O white
dwarfs merge [double-degenerate (DD) scenario]
\cite[e.g.,][]{Arnett1969, Arnett1996, 2000ARA&A..38..191H, Iben1984,
  2012MNRAS.419.1695I, 2014ARA&A..52..107M, Nomoto1982a, Nomoto1994,
  Nomoto1997, 2000AIPC..522...35N, Nomoto2009, Webbink1984}.

Here we focus on the possible evolutionary paths for the accreting
white dwarf to increase its mass to the Chandrasekhar mass in the
binary systems.

\section{Hydrogen Shell-Burning in Accreting White Dwarfs}
\label{sec:1}
            
\subsection {Effects of Mass Accretion on White Dwarfs}
\label{sec:accretion}

     Isolated white dwarfs are simply cooling stars that eventually end 
up as invisible frigid stars.  The white dwarf in a close binary 
system evolves differently because the companion star expands and 
transfers matter over to the white dwarf at a certain stage of its 
evolution.  The mass accretion can {\sl rejuvenate} the cold white 
dwarf \cite[e.g.,][]{Nomoto1977}, which could lead to a SNe Ia 
or accretion-induced collapse (AIC) in some cases.   

     The scenario that possibly brings a close binary system to a SN Ia 
or AIC is as follows: Initially the close binary system consists of two 
intermediate mass stars ($M$ \ltsim 8 $M_{\odot}$).  As a result of 
Roche lobe overflow, the primary star of this system becomes a white 
dwarf composed of carbon and oxygen (C+O).  When the secondary star 
evolves, it begins to transfer hydrogen-rich matter over to the white 
dwarf.  

The mass accretion onto the white dwarf releases gravitational energy
at the white dwarf surface.  Most of the released energy is radiated
away from the shocked region as UV and does not contribute much to
heating the white dwarf interior.  The continuing accretion compresses
the previously accreted matter and releases gravitational energy in
the interior.  A part of this energy is transported to the surface and
radiated away from the surface ({\sl radiative cooling}) but the rest
goes into thermal energy of the interior matter ({\sl compressional
  heating}).  Thus the interior temperature of the white dwarf is
determined by the competition between compressional heating and
radiative cooling; i.e., the white dwarf is hotter if the mass
accretion rate $\dot M$ is larger, and vice versa
\cite[e.g.,][]{Nomoto1982a,Nomoto1982b}.

\subsection{Hydrogen Shell-Flashes}
\label{sec:hydro_burning}
            
\begin{figure}
\begin{center}
\includegraphics[scale=0.45]{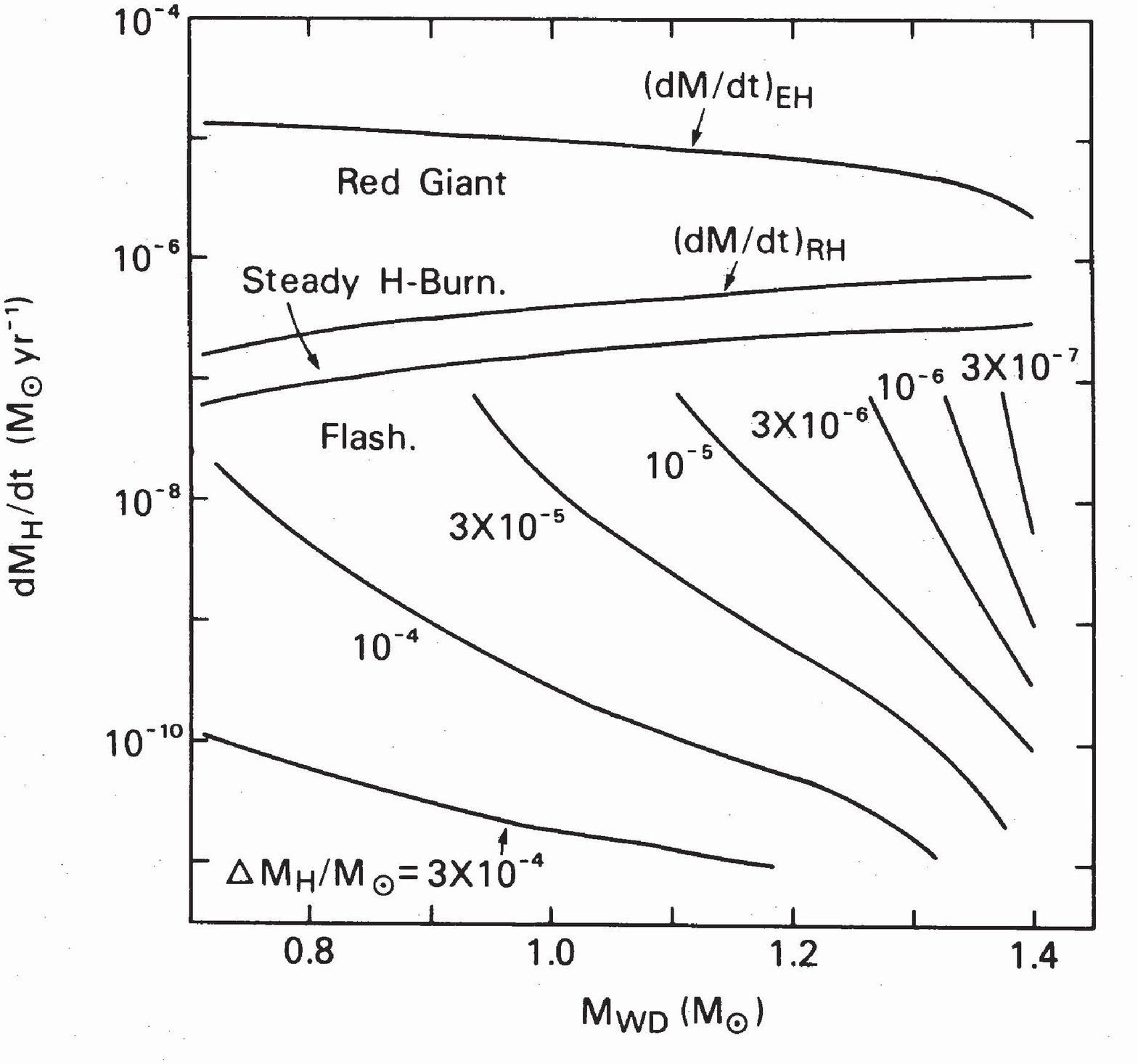}
\end{center}
\caption{
The properties of accreted hydrogen-rich materials as functions of
$M_{{\rm WD}}$ and $dM_{\rm H}/dt$ \cite[]{Nomoto1982a}.  Hydrogen
burning is stable in the region indicated by ``Steady H-Burn'' between
the two lines of $\dot M_{\rm stable}$ in Equation \ref{stableH} and
$\dot M_{\rm cr}$ (=$(dM/dt)_{\rm RH}$) in Equation \ref{steadyH}
\cite[]{kato14}.  In the region below $\dot{M}_{\rm stable}$, hydrogen
shell burning is thermally unstable, and the WD experiences shell
flashes.  Black solid lines indicate the hydrogen-ignition masses
$\Delta M_{\rm H}$, the values of which are shown beside each line.
In the region above $(dM/dt)_{\rm RH}$ (and below the Eddington limit
$(dM/dt)_{\rm EH}$), optically thick winds are accelerated, which
prevents the formation of a red-giant size envelope with the piled-up
accreted material as seen in Fig. \ref{radius} if no wind is included in the
calculation \citep{nom79}. It has been found that optically thick 
wind are accelerated in this region, which prevents the formation 
of a red-giant \cite[]{hac96}.
}
\label{mcmdot}
\end{figure}

Hydrogen shell-burning is ignited when the mass of the accumulated
hydrogen-rich matter reaches the ignition mass $M_{\rm ig} (=\Delta
M_{\rm H})$.  When $M_{\rm ig}$ is reached, the compressional heating
due to accretion is just balanced with the cooling due to heat
conduction \cite[]{nom79,Nomoto1982a}.
$M_{\rm ig}$ is presented as contours on the
$M_{\rm WD}-\dot{M} (=dM_{\rm H}/dt)$ plane in Figure \ref{mcmdot}.
For a given $\dot M$, $M_{\rm ig}$ is smaller for a larger $M_{\rm
  WD}$ because of the smaller radius $R$ and thus higher pressure for
the same mass of accreted matter (see Equation~\ref{pressure_q}
below).  For a given $M_{\rm WD}$, $M_{\rm ig}$ is smaller for a
higher $\dot M$ because of the faster compressional heating and thus
higher temperature of accreted matter.

The stability of the hydrogen burning shell in the accreting white
dwarf is crucial for its evolution.  Figure~\ref{mcmdot} summarizes
the properties of hydrogen shell burning
\cite[]{Nomoto1982a,nom07,kato14}.

\begin{figure}
\begin{center}
\includegraphics[scale=0.37]{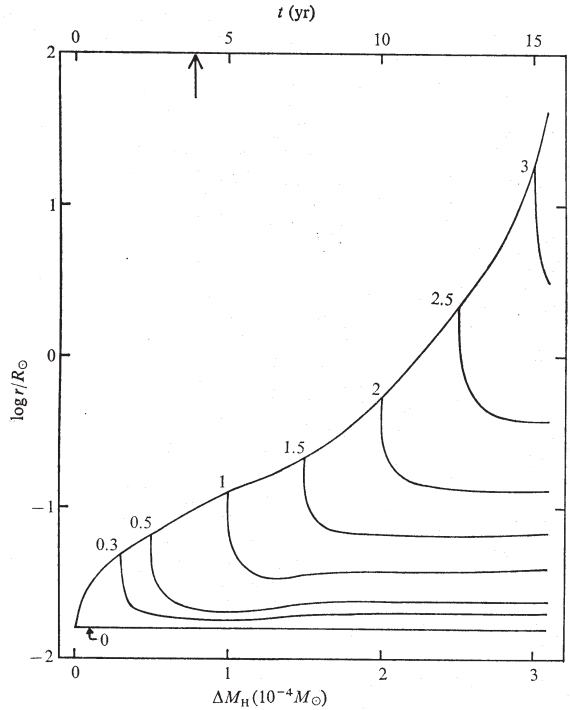}
\end{center}
\caption{
Increase in the radius of the accreting white dwarf upon
the rapid accretion \cite[]{nom79}.
}
\label{radius}
\end{figure}

\noindent
(1) The hydrogen shell burning is unstable to flash in the area below
the solid line to show $\dot M_{\rm stable}$.  This stability line
(dashed line) is approximately represented by \cite{kato14}
\begin{equation}
 \dot M_{{\rm stable}}  = 4.17 \times 10^{-7} \left( {M_{{\rm WD}}
\over M_{{\odot}}} -0.53 \right) ~M_{{\odot}}~{{\rm yr}}^{-1} .
\label{stableH}
\end{equation}

\noindent
(2) Above the dash-dotted line for $\dot M_{\rm cr}$ (= $(dM/dt)_{\rm
  RH}$), the accreted matter is accumulated faster than consumed into
He by H-shell burning.
This critical accretion rate is represented as \cite{kato14}
\begin{equation}
 \dot M_{\rm cr}  = 8.18 \times 10^{-7} \left( {M_{\rm WD} 
\over M_{{\odot}}} -0.48 \right) ~M_{{\odot}}~{\rm yr}^{-1} .
\label{steadyH}
\end{equation}

\noindent
(3) For the region with $\dot M > \dot M_{\rm cr}$, the accreted
matter is piled up to form a red-giant size envelope \cite[]{nom79}.
This could lead to the formation of a common envelope and prevent
further mass accretion onto the white dwarf.  This problem for has
been resolved by the strong optically thick winds \cite[]{hac96, hac99a,
  hac99b}.  If the wind is sufficiently strong, the white dwarf radius
stays small enough to avoid the formation of a common envelope.  Then
steady hydrogen burning increases its mass at a rate $\dot M_{\rm cr}$
by blowing the extra mass away in a wind.

\noindent
(4) In the area $\dot M_{\rm stable} < \dot M < \dot M_{\rm cr}$,
accreting white dwarfs are thermally stable so that hydrogen burns
steadily in the burning shell. Then the white dwarf mass increases at
a rate of $\dot M$.

\noindent
(5) For $\dot M < \dot M_{\rm stable}$, the flash of hydrogen shell
burning is stronger (weaker) for lower (higher) $\dot M$ and thus for
larger (smaller) $M_{{\rm ig}}$ and larger (smaller) $M_{\rm WD}$.

\begin{figure}
\begin{center}
\includegraphics[scale=0.37]{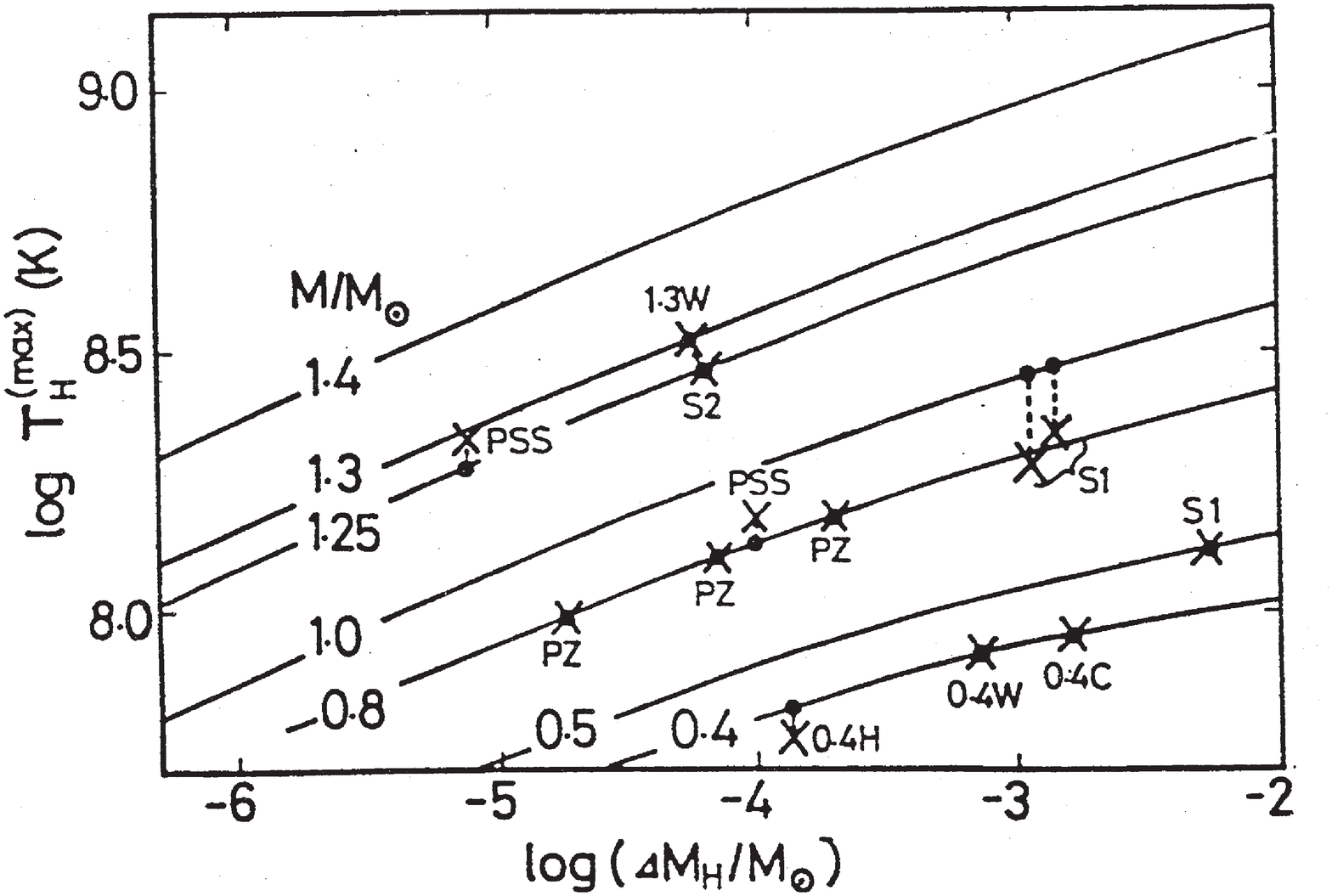}
\end{center}
\caption{
Maximum temperature attained during the hydrogen shell
flash as a function of the accreted mass \mdel{H} and the white dwarf
mass $M$.  Analytical values (solid curves) are compared with those
obtained by hydrodynamical calculations (X-mark) (see \cite[]{Sugimoto1979}
for details).
}
\label{peakT}
\end{figure}

The progress and the strength of the flashes are determined by 
two parameters ($P^*, \Omega^*$), 
\begin{equation} 
  P^* = {GM_{\rm WD} M_{\rm ig} \over 4 \pi R^4},
  \qquad \Omega^* = {GM\over R},
\label{pressure_q}
\end{equation}
\noindent
i.e., by the pressure and potential at the burning shell corresponding
to completely {\sl flat} configuration
\cite[]{Sugimoto1978,Sugimoto1979}.  A set of ($P^*, \Omega^*$) can be
transformed into ($M, M_{\rm ig}$) since the radius at the burning
shell is well approximated by the WD radius $R$ which is smaller for
larger $M_{\rm WD}$.

     Progress of the shell flash can be treated semi-analytically.  
Initially the temperature  at the burning shell increases along $P = 
P^*$.   As the nuclear energy is released, the pressure decreases as a 
result of expansion, which is described by $P = fP^*$.  Here the 
flatness parameter $f$ is unity for plane-parallel configuration and
$f <$ 1 for more spherical configuration.  This is expressed as 
\begin{equation} 
{1\over f(V, N)} = \sum_{k=0}^{\infty} b_k, \qquad b_0 = 1, \quad b_k 
= b_{k-1} {k+3\over N+k+1}{N+1\over V}, 
\qquad  V = r/H_{\rm p}, 
\label{flatness}
\end{equation}
\noindent
where $N$ and $H_{\rm p}$ denote the polytropic index for the 
convective envelope and the scale height of pressure, respectively 
\cite[]{Sugimoto1978}.  As the specific entropy $s$ in the 
hydrogen-burning shell increases, $f$ decreases because of increasing 
$H_{\rm p}$, i.e., expansion of the accreted envelope. This 
corresponds to the change in the configuration of the burning shell 
from plane parallel to spherical \citep{Sugimoto1980}.

Then the temperature reaches its maximum $T_{\rm H}^{\rm max}$, which
is higher for higher $P^*$ and thus for higher $M$ (smaller $r$) and
larger \mdel{H} (Eq. \ref{pressure_q}).  Such a relation between
$T_{\rm H}^{\rm max}$ and \mdel{H} for several $M$ obtained from
Equation \ref{flatness} is shown in Figure \ref{peakT}
\cite[]{Sugimoto1979}.  Results of some hydrodynamical calculations
(X-mark) are in excellent agreement with the corresponding analytical
predictions (filled circles).  (Some discrepancies are likely to be
caused by a coarse zoning in numerical calculations).

The results in Figures \ref{mcmdot} and \ref{peakT} show that
generally smaller $M$ and higher $\dot M$ lead to a weaker flash
because of the lower pressure at the flashing shell.  For $\dot M
\sim$ 2 \e{-7} -- \ee{-8} $M_{\odot}$ yr$^{-1}$, the flash is so weak
that the ejected mass is smaller than the mass which hydrogen burning
converts into helium.  Then the mass of the He layer can increase.

For $\dot{M} \sim 10^{-9} - 10^{-8} M_\odot~{\rm yr}^{-1}$, the
H-flash is stronger so that the mass of the He layer grows but at much
lower rate.

For slow accretion ($\dot M$ \ltsim 1 \e{-9} $M_\odot$ yr$^{-1}$),
hydrogen shell flash is strong enough to grow into a {\sl nova}
explosion, which leads to the ejection of most of the accreted matter
from the white dwarf \cite[e.g.,][]{Nariai1980}.
Moreover, a part of the original white dwarf matter is
dredged up and lost in the outburst wind.  Then $M_{{\rm WD}}$
decreases after the nova outburst.  For these cases, the white dwarf
does not become a {\sl supernova} since its mass hardly grows.
However, if the white dwarfs are close to the Chandrasekhar mass,
novae could grow into AIC of SN Ia because the ejected mass from nova
explosion is found to be significantly smaller than the accreted mass
\cite[]{Starrfield1991}.

In these cases, the hydrogen flash recurs, and the recurrence period
is proportional to $M_{{\rm ig}}$/$\dot{M}$, which is shorter for
higher $\dot{M}$.  If the recurrence period is short, the flashes are
observed as {\sl recurrent novae}, which occurs in the upper-right region of
Figure \ref{mcmdot}.

\begin{figure}
\begin{center}
\includegraphics[scale=0.37]{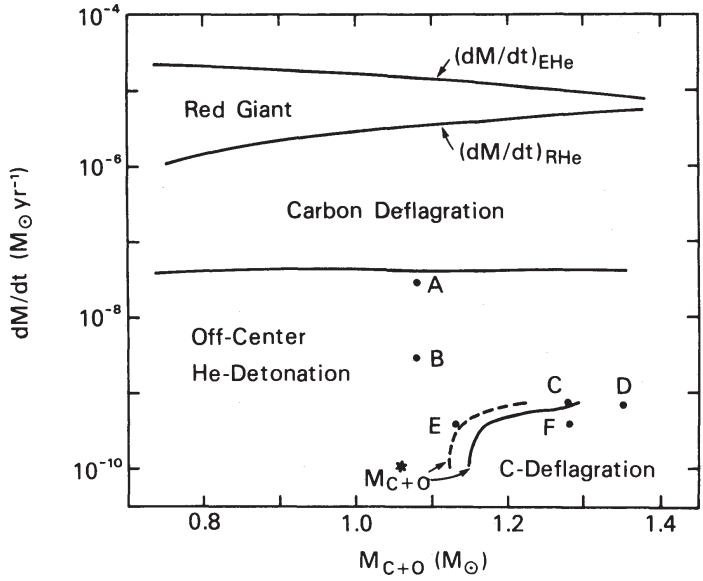}
\end{center}
\caption{
The properties of He accreting white dwarf as functions of
$M_{\rm WD}$ and $dM/dt$ \cite[]{Nomoto1982a}.  In the region above
$(dM/dt)_{\rm RHe}$ (and below the Eddington limit
$(dM/dt)_{\rm EHe}$), the accreted He envelope is extended to a
red-giant size.  In the region below $\dot{M}_{\rm RHe}$, He shell
burning is thermally unstable, and the WD experiences shell flashes.
}
\label{HeAccret}
\end{figure}

\begin{figure}
\begin{center}
\includegraphics[scale=0.45]{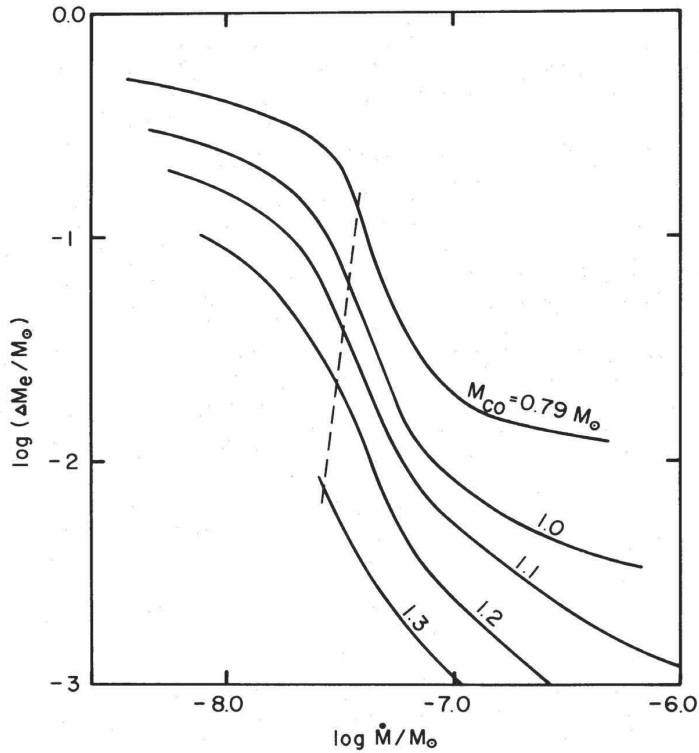}
\end{center}
\caption{
The mass of accreted He envelope, \mdelp{e}, at the
He ignition as a function of $\dot M$  and the mass of underlying 
C+O  core, $M_{\rm CO}$ \cite[]{Kawai1987}.
}
\label{Heigmass}
\end{figure}

\begin{figure}
\begin{center}
\includegraphics[scale=0.45]{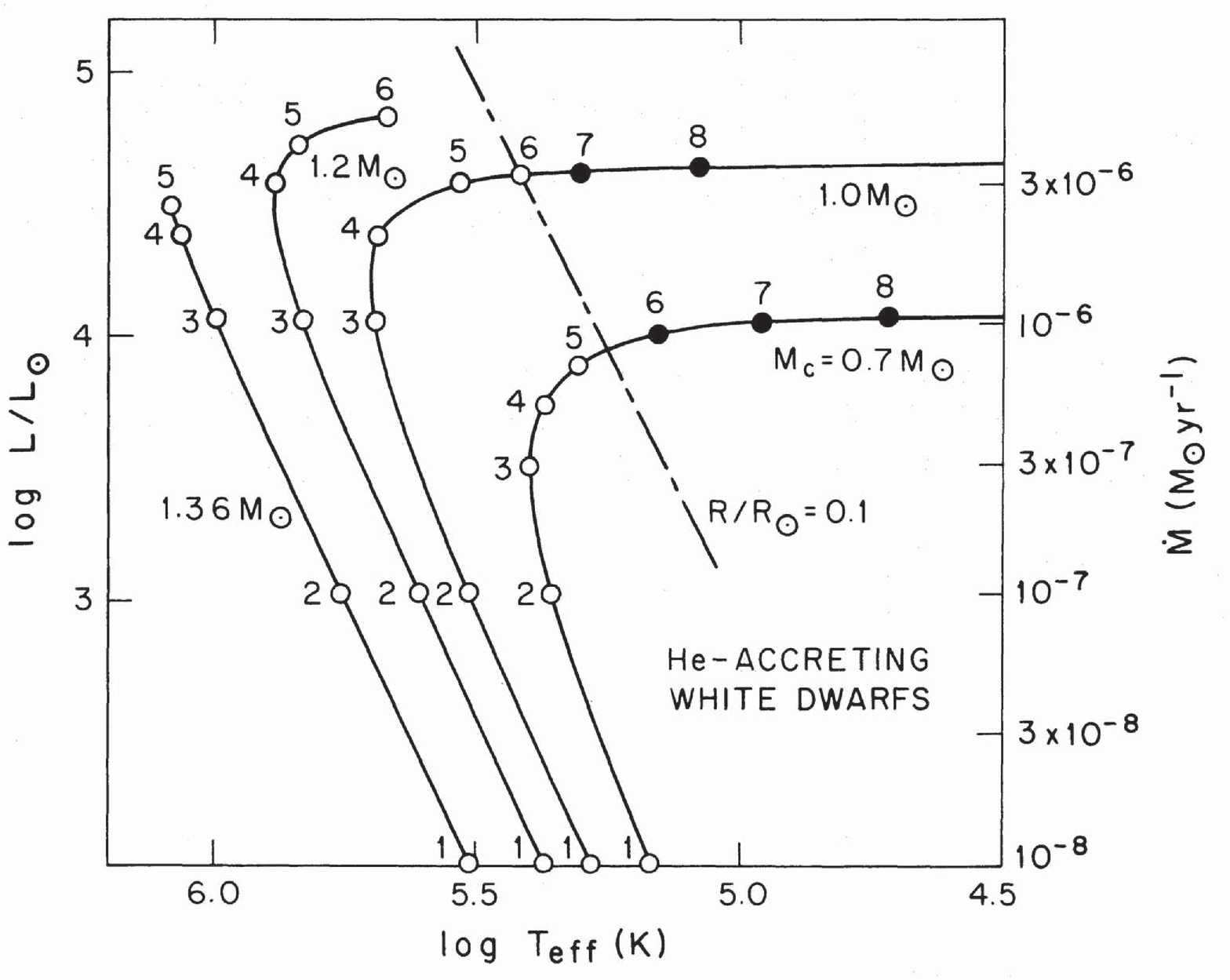}
\end{center}
\caption{
Locations of He-accreting steady state white dwarf models in the H-R
diagram.  The accretion rate is indicated on the right vertical axis.
The solid lines connect models with the same C+O core mass.  The
thermally unstable and stable models are indicated by open and filled
circles, respectively \cite[]{Kawai1988}.
}
\label{HeHR}
\end{figure}

\section {Evolution of Helium Accreting White Dwarfs}

\subsection{Accretion of Helium}

As discussed above, a thin He layer is produced and grows by H-burning
for $\dot{M} > 10^{-9} M_\odot~{\rm yr}^{-1}$.  The He layer grows
also by direct transfer of helium if the companion is a He star
\cite[e.g.,][]{Iben1987}.
Further evolution and final fates depend on the accretion rate of He
and the mass of the C+O core \mm{CO} as summarized in Figure
\ref{HeAccret}.

When a certain mass \mdel{He} is accumulated, He shell-burning is
ignited; the solid lines in Figure \ref{Heigmass} show \mdel{He} as a
function of $\dot M$ and the white dwarf masses $M$
\cite[]{Kawai1987}.  It is seen that \mdel{He} is larger for the
slower mass-accumulation rate of the He layer $\dot{M_{\rm He}}$.

If $(dM/dt)_{\rm EHe} \gsim \dot M \gsim (dM/dt)_{\rm RHe}$ as in
Figure \ref{HeAccret}, the accreted He envelope is extended to a
red-giant size as seen in Figure \ref{HeHR}.  In the region below
$\dot{M}_{\rm RHe}$, He shell burning is thermally unstable, and the
WD experiences shell flashes as discussed in the next subsection.

If the accretion of He is as slow as $\dot M$ \ltsim 1 \e{-9}
$M_{\odot}$ yr$^{-1}$, the accreted material is too cold to ignite He
burning, so that the white dwarf mass increases.  An exception is the
case with \mm{CO} \ltsim 1.1 $M_{\odot}$ where pycnonuclear He burning
is ignited at high enough densities \cite[]{Nomoto1982a,Nomoto1982b}.

\begin{figure}
\begin{center}
\includegraphics[scale=0.40]{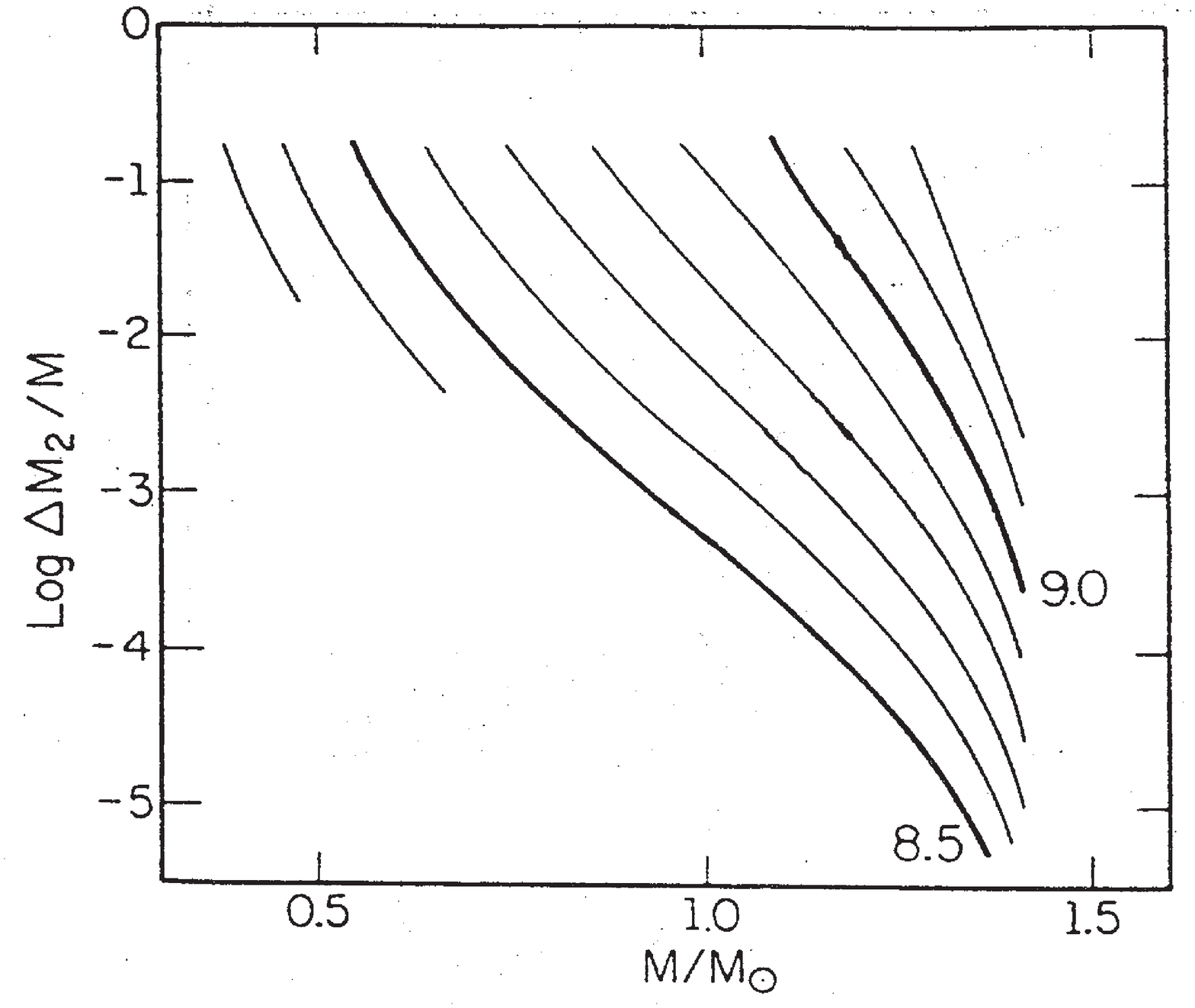}
\end{center}
\caption{
Maximum temperature log $T$ (K) attained during the
He shell flash as a function of the accreted mass \mdel{2} and the
white  dwarf mass $M$ \cite[]{Fujimoto1982}.
}
\label{HeTpeak}
\end{figure}

\subsection {Helium Shell-Flashes and Detonation}
\label{sec:he_flash}

In the early stages of He shell-burning, the He envelope is
electron-degenerate and geometrically almost flat.  Because of the
almost constant pressure at the bottom of the He-burning shell
(Eq. \ref{pressure_q}), the temperature there increases and makes a He
flash.  Heated by He burning, the He envelope gradually expands, which
decreases the pressure.  Then, the temperature attains its maximum and
starts decreasing.

The maximum temperature attained during the He shell-flash depends
on \mdel{He} and $M$ (Figure \ref{HeTpeak}: \cite{Fujimoto1982}) as
discussed for hydrogen in \S \ref{sec:hydro_burning}.  The maximum
temperature is higher for more massive WD and more massive envelope
because of higher pressure.  The strength depends on the He envelope
mass $M_{{\rm env}}$, thus depending mainly on $\dot M$ as follows.

\subsubsection{He Detonation}

For \mdp{det} \gtsim $\dot M$ \gtsim \ee{-9} $M_{\odot}$ yr$^{-1}$,
the He shell-flash is strong enough to initiate an off-center He
detonation, which prevents the white dwarf mass from growing
\cite[e.g.,][]{Nomoto1982b,Woosley1986}.  Here we adopt \mdp{det}
$\sim$ 1 \e{-8} $M_{\odot}$ yr$^{-1}$, since the \nco (NCO) reaction
ignites weak He flashes \cite[]{Hashimoto1986,Limongi1991} if the mass
fraction of CNO elements in the accreting material exceeds 0.005.  For
smaller CNO abundances, the NCO reaction is not effective and thus
\mdp{det} $\sim$ 4 \e{-8} $M_{\odot}$ yr$^{-1}$ \cite[]{Nomoto1982a}.

Two dimensional hydrodynamical simulations after the initiation of He
detonation have been performed by several groups (e.g.,
\cite{Livne1991,Nomoto2017} and references therein).  The outcome is the
off-center He-detonation, which develops into double detonation
supernovae.

\subsubsection{Recurring He Flashes}

For intermediate accretion rates (3 \e{-6} $M_{\odot}$ yr$^{-1}$
\gtsim $\dot M$ \gtsim 1 \e{-8} $M_{\odot}$ yr$^{-1}$), He
shell-flashes are of moderate strength, thereby recurring many times
to increase the white dwarf mass
\cite[]{Taam1980,Fujimoto1982,Nomoto1982b,kato2017}.  When the white
dwarf mass becomes close to the Chandrasekhar mass, either
thermonuclear explosion or collapse would occur.

Nucleosynthesis in such He shell-flashes has been calculated for
various set of ($M_{{\rm WD}}$, $M_{{\rm env}}$) \cite[]{nom13, shen07}.
For higher maximum temperatures, heavier elements, such as $^{28}$Si
and $^{32}$S, are synthesized.  However, the maximum temperature is
not high enough to produce $^{40}$Ca.  After the peak, some amount of
He remains unburned in the flash and burns into C+O during the stable
He shell-burning.  In this way, it is possible that an interesting amount
of intermediate mass elements, including Si and S, already exist in
the unburned C+O layer at $M_r \geq 1.2 M_{\odot}$.

\begin{figure}\begin{center}
\includegraphics[scale=0.37]{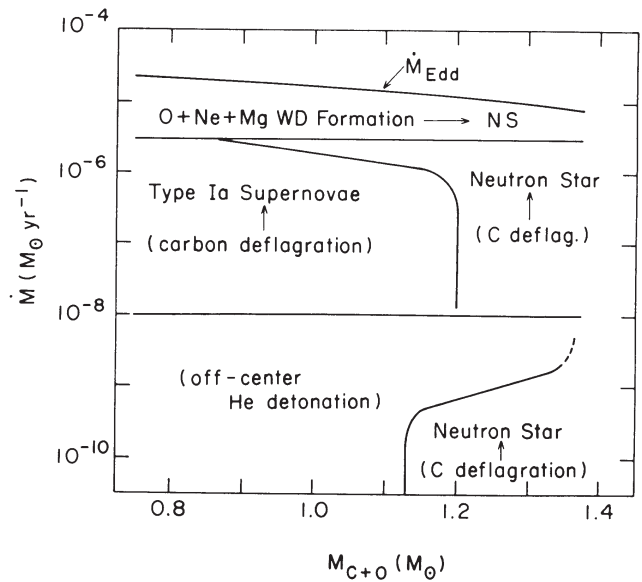}
\end{center}
\caption{
  The final fate of C+O white dwarfs expected for their initial mass
  and accretion rate $\dot M$ of C+O materials \cite[]{Nomoto1991}.
}
\label{mcmHedot}
\end{figure}

\subsection{Neutron Star Formation}

For the C+O white dwarfs, whether they explode or collapse depends not
only on $\dot M$ but also on the initial mass \mmp{CO} as summarized
in Figure \ref{mcmHedot}.  For \mmp{CO} $<$ 1.2 $M_{\odot}$,
substantial heat inflow from the surface layer into the central region
ignites carbon at relatively low central density ($\rho_{\rm c}$
$\sim$ 3 \e{9} g cm$^{-3}$) \cite[]{Nomoto1984}.

On the other hand, if the white dwarf is sufficiently massive and cold
at the onset of accretion, the central region is compressed only
adiabatically, thereby being cold (and solid) when carbon is ignited
in the center of density as high as \ee{10} g cm$^{-3}$.  C-burning at
such high densities is likely to result in a collapse due to rapid
electron capture in NSE (nuclear statistical equilibrium).

Neutron star formation is a possible outcome from high C-accretion
rates as discussed in the next section on merging C+O WD (as
indicated in Figure \ref{mcmHedot}).

\begin{figure}\begin{center}
\includegraphics[scale=0.37]{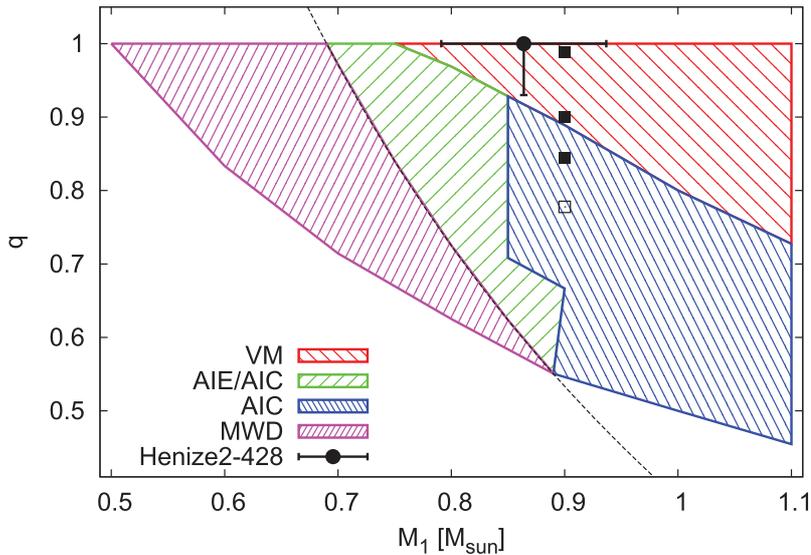}
\end{center}
\caption{
  Final outcomes of our merger simulations in the mass ratio versus
  total mass diagram, i.e., violent merger, AIC (accretion-induced
  collapse), and AIE (accretion-induced explosion)/AIC
  \cite[]{Sato2017}.  The black circle shows the possible ranges of 
  the mass ratio and total mass of the central system of Henize 2-428 
  \cite[]{Sato2016}. The black squares denote the models calculated
  in \cite{Pakmor2011}.  Filled squares indicate the models which
  satisfied the detonation condition of \cite{Seitenzahl09}, while the
  open square indicates the model that does not.
}
\label{merge}
\end{figure}

\section {Merging of Double C+O White Dwarfs}

Merging of double C+O white dwarfs is estimated to take place as
frequently as SNe Ia \cite[]{Iben1984,Webbink1984}.  \cite{Sato2017}
performed SPH simulations of merging and summarized the final outcomes
of mergers in the primary white dwarf mass ($M_1$) versus the mass
ratio of the two white dwarfs ($q = M_2/M_1$) in Figure \ref{merge} as
follows:
\noindent
(1) Violent Merger (VM): If the temperature of the shock-heated merged
region becomes high enough, i.e., the timescale of the temperature
rise due to C-burning becomes shorter than the dynamical timescale,
C-detonation is generated which then triggers the central C-detonation
at a relatively low central density as determined by $M_1$.

In Figure \ref{merge} \cite[]{Sato2017}, the black squares denote the
models calculated in \cite{Pakmor2011}.  Filled squares indicate the
models which satisfied the detonation condition of
\cite{Seitenzahl09}, while the open square indicates the model that
does not.
\noindent
(2) Accretion-induced collapse (AIC): If the temperature at the merged
region leads to a stable off-center C-burning, C-flame propagates
through the center to convert the C+O white dwarf into the O+Ne+Mg
white dwarf, which would eventually collapse due to electron capture.

Such a C-ignition and subsequent flame propagation have been
approximately simulated by spherical models. After the smaller mass
white dwarf fills its Roche lobe, mass transfer of carbon onto the
more massive white dwarf would be very rapid, which ignites off-center
C-burning if $\dot M$ \gtsim 2.7 \e{-6} $M_{\odot}$ yr$^{-1}$
\cite[]{Nomoto1985}.  Subsequent flame propagation converts entire C+O
into O+Ne+Mg \cite[]{Saio2004}, which leads to collapse as induced by
electron capture.

\noindent
(3) Accretion-induced explosion (AIE): If the temperature after merging
is too low to ignite carbon, the C+O white dwarf would increase its
mass toward the Chandrasekhar mass (unless the accretion ignites
off-center C burning) and leads to an SN Ia \cite[]{Yoon2007}.

\noindent
(4) C+O white dwarf (MWD): If the total mass $M_{\rm tot}= M_1 + M_2$
of the two white dwarfs does not exceed the Chandrasekhar mass, the
above cases (2) and (3) could not occur.  Instead, the merger results
in the formation of a single C+O white dwarf.


\section{Four Cases of Pre-Explosion Configurations of Type Ia Supernovae} 

\begin{figure}
\begin{center}
\includegraphics[scale=0.6]{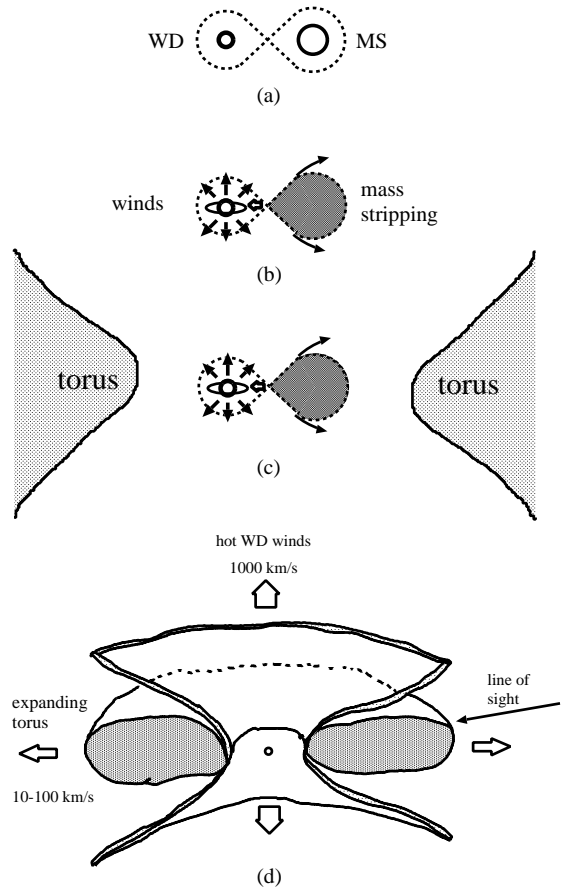}
\end{center}
\caption{
A schematic configuration of a binary evolution including
mass-stripping effect \cite[]{Hachisu2008a}.  (a) Here we start a pair
of a C+O WD and a more massive main-sequence (MS) star with a
separation of several to a few tens of solar radii.  (b) When the
secondary evolves to fill its Roche lobe, mass transfer onto the WD
begins.  The mass transfer rate exceeds a critical rate $M_{\rm cr}$
for optically thick winds.  Strong winds blow from the WD.  (c) The
hot wind from the WD hits the secondary and strips off its surface.
(d) Such stripped-off material forms a massive circumstellar disk or
torus and it gradually expands with an outward velocity of $\sim
10-100$~km~s$^{-1}$.  The interaction between the WD wind and the
circumstellar torus forms an hourglass structure.  The WD mass
increases up to $M_{{\rm Ia}}= 1.38 ~M_{\odot}$ and explodes as an SN
Ia.
}
\label{stripping_evolution}
\end{figure}

Based on the above properties of accretion-induced hydrogen shell
burning, the binary system in the SD scenario evolves through stages
(a)-(d) below (also shown in Figures~\ref{stripping_evolution}a--d)
\cite[]{Hachisu2008a}.

The more massive (primary) component of a binary evolves to
a red giant star (with a helium core) or an AGB star (with a 
C+O core) and fills its Roche lobe.
Mass transfer from the primary to the secondary begins and a common
envelope is formed.  After the first common envelope evolution,
the separation shrinks and the primary component
becomes a helium star or a C+O WD.  The helium star
evolves to a C+O WD after a large part of helium is exhausted
by core-helium-burning.
We eventually have a close pair of a C+O WD and a main-sequence
(MS) star (Figure \ref{stripping_evolution}a).

Further evolution of the system depends on the binary parameters.
Depending on at which stage SNe Ia are triggered, the SD scenario
predicts the following four variations of SNe Ia.

\subsection{SNe Ia - Circumstellar Matter (CSM)} 
\index{circumstellar matter} 

After the secondary evolves to fill its Roche lobe, the mass transfer
to the WD begins.  This mass transfer occurs on a thermal timescale
because the secondary mass is more massive than the WD.  The mass
transfer rate exceeds $\dot M_{\rm cr}$ for the optically thick wind
to blow from the WD \cite[]{hac96, hac99a, hac99b} (Figure
\ref{stripping_evolution}b).

Optically thick winds from the WD collide with the secondary
surface and strip off its surface layer.  This mass-stripping
attenuates the rate of mass transfer from the secondary to the WD,
thus preventing the formation of a common envelope for a more massive
secondary in the case with than in the case without this effect.  Thus
the mass-stripping effect widens the donor mass range of SN Ia
progenitors (Figure \ref{stripping_evolution}c).

Such stripped-off matter forms a massive circumstellar torus on
the orbital plane, which may be gradually expanding with an outward
velocity of $\sim 10-100$~km~s$^{-1}$ (Fig.
\ref{stripping_evolution}d), because the escape velocity from the
secondary surface to L3 point is $v_{\rm esc}\sim 100$ ~km~s$^{-1}$.
Subsequent interaction between the fast wind from the WD and the very
slowly expanding circumbinary torus forms an hourglass structure
(Fig. \ref{stripping_evolution}c--d).  When we observe the SN Ia from
a high inclination angle such as denoted by ``line of sight,''
circumstellar matter can be detected as absorption lines like in SN
2006X.

This scenario predicts the presence of several types of circumstellar
matter around the binary system, which are characterized various wind
velocities $v_{{\rm w}}$: (1) white dwarf winds with such high
velocities as $v_{{\rm w}} \sim 1000$ km s$^{-1}$, (2) slow dense matter
stripped off the companion star by the white dwarf wind, (3) slow wind
matter ejected from a red-giant, and (4) moderate wind velocities
blown from the main-sequence star.

\begin{figure}
\begin{center}
\includegraphics[scale=0.7]{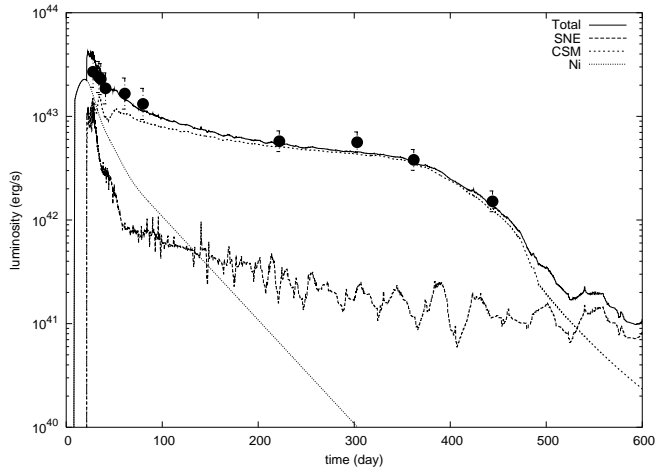}
\end{center}
\caption{The observed light curve of SN Ia 2002ic (filled circles) and
  the calculated light curve with circumstellar
  interaction \cite[]{nom05}.}
\label{sn2002ic}
\end{figure}

The above features are supported by observations of the presence of
circumstellar matter in some SNe Ia \cite[]{2007Sci...317..924P,
2011Sci...333..856S, 2012ApJ...752..101F}, 
and the detection of H in
circumstellar-interaction type supernovae (Ia/IIn) such as SN 2002ic
\cite[]{2003Natur.424..651H}. 
SN 2002ic shows the typical spectral features of SNe Ia near maximum
light, but also apparent hydrogen features that have been absent in
ordinary SNe Ia.  Its light curve has been reproduced by the model of
interaction between the SN Ia ejecta and the H-rich circumstellar
medium (Fig. \ref{sn2002ic}) \cite[]{nom05}.

\subsection{SNe Ia - Supersoft X-ray Sources}
\index{supersoft X-Ray source}

When the mass transfer rate decreases to the following range: $\dot
M_{\rm stable} < \dot M < \dot M_{\rm cr}$, optically thick winds
stop, and the WDs undergo steady H-burning.  The WDs are observed as
supersoft X-ray sources (SSXSs) until the SN Ia explosion.  The
stripped-off material forms circumstellar matter (CSM) but it has been
dispersed too far to be detected immediately after the SN Ia
explosion.

\subsection{SNe Ia - Recurrent Novae}
\index{recurrent novae}

When the mass transfer rate from the secondary further decreases below
the lowest rate of steady hydrogen burning, i.e., ${\dot M}_{\rm
  transfer} < {\dot M}_{\rm stable}$, hydrogen shell burning is
unstable to trigger a mild flashes, which recur many times in a short
period as a recurrent nova (RN) \cite[e.g.,][]{nom07}.  Its recurrent
period is as short as $\sim 1$ yr, which can be realized for high $M$
and high $\dot M$ as discussed in \S \ref{sec:hydro_burning}.  These
flashes burn a large enough fraction of accreted hydrogen to increase
$M$ to SNe Ia.

Observationally, PTF11kx \cite[]{2012Sci...337..942D}
provides strong evidences that the accreting white dwarf was
a recurrent nova and the companion star was a red supergiant.

\subsection{SNe Ia - He White Dwarf Remnants}
\index{He white dwarfs}

In the rotating white dwarf scenario, which will be discussed in a
later section \cite[e.g.,][]{2015ApJ...805...L6}, ignition of central
carbon burning is delayed in some cases due to the larger
Chandrasekhar mass of the rotating white dwarfs than non-rotating
white dwarfs.  This delay time after the end of accretion up to the SN
Ia explosion depends on the time scale of angular momentum loss from
the C+O white dwarfs, and could be long enough for the companion star
to evolve into a He white dwarf and for circumstellar materials to
disperse.  For such a delayed SN Ia, it would be difficult to detect a
companion star or circumstellar matter.

\subsection{Companion Stars in the SD Scenario}

\begin{figure}
\begin{center}
\includegraphics[scale=0.55]{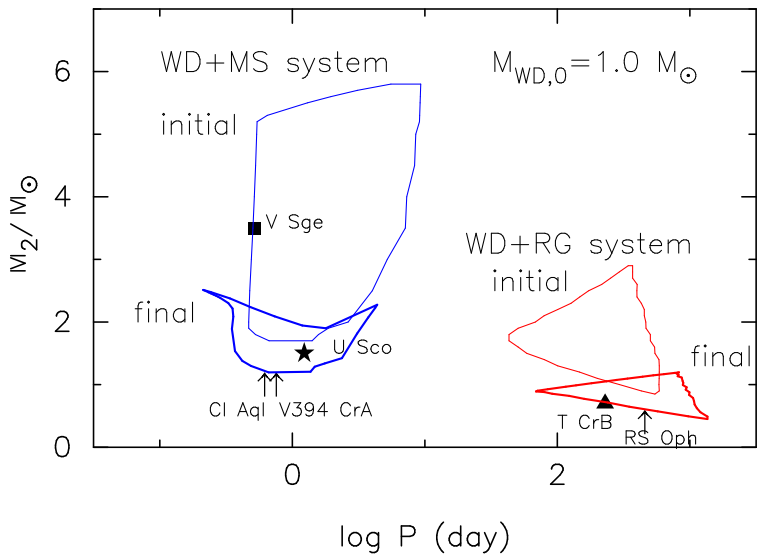}
\end{center}
\caption{
The regions that produce SNe Ia are plotted in the $\log P - M_2$
(orbital period -- secondary mass) plane for the (WD+MS) system
({\it left}) and the (WD+RG) system ({\it right}) \cite[]{hacDtd08}.
Currently known positions of the recurrent novae and supersoft X-ray
sources are indicated by a star mark ($\star$) for U Sco, a triangle
for T CrB, a square for V Sge, but by arrows for the other three
recurrent novae, V394 CrA, CI Aql, and RS Oph.  Two subclasses of the
recurrent novae, the U Sco type and the RS Oph type, correspond to the
WD + MS channel and the WD + RG channel of SNe Ia, respectively.
\label{m2p}}
\end{figure}

In SD scenario, SNe Ia can occur for a wide range of $\dot{M}$. The
progenitor white dwarfs can grow their masses to the Chandrasekhar
mass by accreting hydrogen-rich matter at a rate as high as $\dot{M}
\gtrsim 10^{-7} - 10^{-6} M_{{\odot}}$~yr$^{-1}$
\cite[e.g.,][]{hac96, li97, hac99a, hac99b, 2000A&A...362.1046L, han04, nom00}.

Two types of binary systems can provide such high accretion rates,
i.e., (1) a white dwarf and a lobe-filling, more massive (up to $\sim
7 M_\odot$), slightly evolved main-sequence or sub-giant star (WD+MS),
and (2) a white dwarf and a lobe-filling, less massive (typically
$\sim 1 M_{\odot}$), red-giant (WD+RG) \cite[]{hac99a, hac99b}.  Figure
\ref{m2p} shows these two regions of (WD+MS) and (WD+RG) in the $\log
P - M_2$ (orbital period -- secondary mass) plane \cite[]{hacDtd08}.

Here the metallicity of $Z=0.02$ and the initial white dwarf mass of
$M_{\rm WD, 0}= 1.0 ~M_{\odot}$ are assumed.  The initial system inside
the region encircled by a thin solid line (labeled ``initial'')
increases its WD mass up to the critical mass ($M_{\rm Ia}= 1.38
M_{\odot}$) for the SN Ia explosion, the regions of which are encircled
by a thick solid line (labeled ``final'').

Note that the ``initial'' region of WD + MS systems extends up to such
a massive ($M_{2,0} \sim 5-6 ~M_{\odot}$) secondary, which consists of a
very young population of SNe Ia with such a short delay time as $t
\lesssim 0.1$~Gyr.  On the other hand, the WD + RG systems with a less
massive RG ($M_{2,0} \sim 0.9-1.0 ~M_{\odot}$) consist of a very old
population of SNe Ia of $t \gtrsim 10$ Gyr.

The delay time distribution (DTD) of SNe Ia on the basis of the above SD
model (Fig. \ref{m2p}) has a featureless power law in good
agreement with the observation \cite[]{hacDtd08}.  This is because the
mass of the secondary star of the SN Ia system ranges from
$M_{2.0} \sim 0.9$ to $6 ~M_{\odot}$ due to the effects of the WD winds
and the mass stripping.  In our model, moreover, the number ratio of
SNe Ia between the WD + MS component and the WD + RG component is
$r_{\rm MS/RG} = 1.4$.  Such almost equal contributions of the two
components help to yield a featureless power law.

\begin{figure}[ht]
\begin{center}
\includegraphics*[scale=0.4,angle=270]{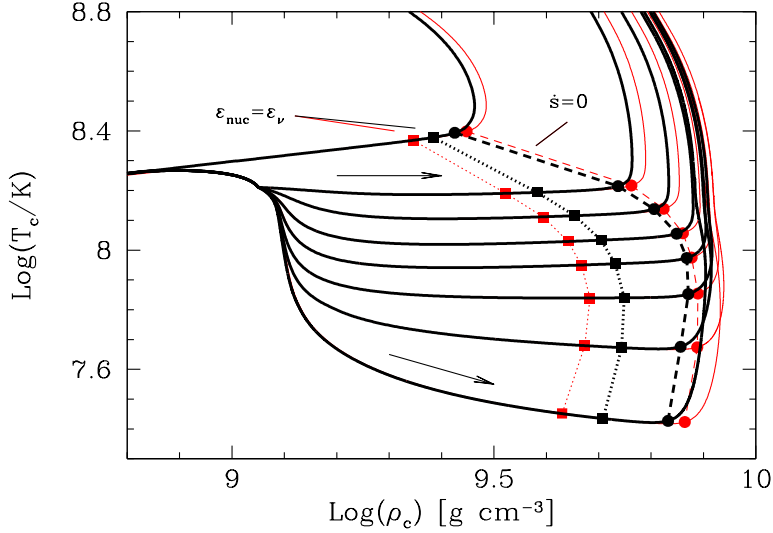}
\end{center}
\caption{The evolutionary tracks of the center
  of the WD up to the onset of the hydrodynamical
  explosion \cite[]{2015ApJ...805...L6}. 
  $\varepsilon_{\rm nuc}=\varepsilon_{\nu}$ indicates the
  conditions at which neutrino losses equal nuclear energy release
  while $\dot{S}=0$ show the stages at which central entropy per
  baryon begins to increase. Thick black and thin red lines correspond
  to the treatments of screening given by \cite{2000ApJ...539..888K}
  and \cite{2012A&A...538A.115P}, respectively. Arrows indicate the
  sense of the evolution.}
\label{fig:center} 
\end{figure}

\subsection{Rotating White Dwarf}
\index{Rotation}

\subsubsection{Uniform Rotation and Delayed Carbon Ignition}

In the above sections, some observations that support the SD scenario
are given.  However, there has been no direct indication of the
presence of companions, e.g., the lack of companion stars in images of
SN 2011fe \cite[]{Li2011}
and some Type Ia supernova remnants \cite[]{2012Natur.481..164S}.
The rotating white dwarf scenario solves this missing-companion problem
\cite[]{2011ApJ...730L..34J, 2011ApJ...738L...1D, 2012ApJ...756L...4H,
  2015ApJ...805...L6}.

The rotating WD evolves as follows \cite[]{2012ApJ...756L...4H,
  2015ApJ...805...L6, pie03}.

\noindent
(1) For certain ranges of binary parameters, the accretion rate
($\dot{M}$) always exceeds $10^{-7} M_{\odot}$ y$^{-1}$ so that the WD
increases its mass until it undergoes ``prompt'' carbon ignition.  The
mass of the uniformly rotating WD at the carbon ignition, $M_{\rm
  ig}^{\rm R}$, is larger for smaller $\dot{M}$.  For $\dot{M} =
10^{-7} M_{\odot}$ y$^{-1}$, $M_{\rm ig}^{\rm R} =$ 1.43~$M_{\odot}$,
which is the largest mass because nova-like hydrogen flashes prevent
the the WD mass from growing for the lower $\dot{M}$.  Because of the
centrifugal force in the rotating WD, $M_{\rm ig}^{\rm R}=$
1.43~$M_{\odot}$ is larger than $M_{\rm ig}^{\rm NR} =$
1.38~$M_{\odot}$ \cite[]{Nomoto1984}.

\noindent
(2) For adjacent ranges of binary parameters, the mass of the rotating
WD exceeds $M_{\rm ig}^{\rm NR} =$ 1.38~$M_{\odot}$ but does not reach
$M_{\rm ig}^{\rm R}=$ 1.43~$M_{\odot}$ because of the decreasing
accretion rate.  After the accretion rate falls off, the WD undergoes
the angular momentum-loss (J-loss) evolution.  The exact mechanism and
the time scale of the J-loss are highly uncertain, although the
magneto-dipole braking WD is responsible.  J-loss induces the
contraction of the WD, which leads to the ``delayed'' carbon ignition
after the ``delay'' time due to neutrino and radiative cooling.

Figure~\ref{fig:center} shows the evolution of the center of the WD
since before the end of accretion up to the onset of the
hydrodynamical stage \cite[]{2015ApJ...805...L6}.  The upper-most line
corresponds to the case (1) evolution that leads to the ``prompt''
carbon ignition. Below that, the lines from upper to lower correspond
to the evolutions with increasing J-loss timescale ($\tau_{\rm J}=$~1,
3, 10, 30, 100, 300, and 1000~Myr, respectively), that lead to the
``delayed'' carbon ignition.

\begin{figure}[ht]
\begin{center}
\includegraphics*[scale=0.85]{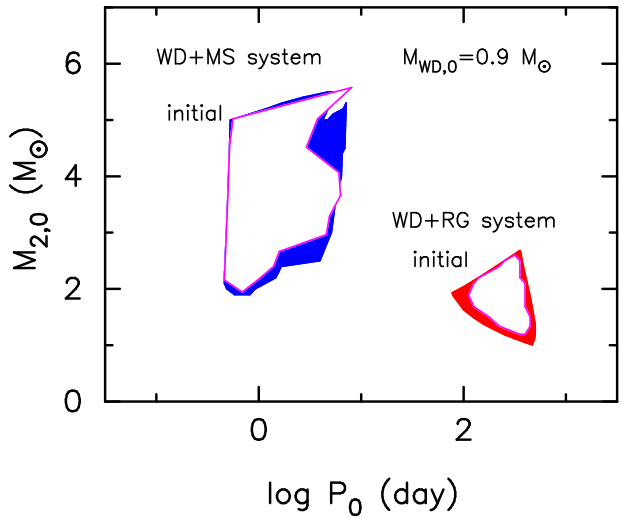}
\end{center}
\caption{
The outcome of the binary evolution of the WD + companion star systems
is shown in the parameter space of the initial orbital period $P$ and
the companion mass $M_2$ for the initial WD mass of $0.9 M_{\odot}$
\cite[]{2015ApJ...805...L6}.  The mass $M$ of the WD starting from the
``painted'' region reaches $1.38~M_{\odot} < M < 1.43~M_{\odot}$
(delayed carbon ignition), while the systems starting from the blank
region encircled by the solid line reach $M = 1.43~M_{\odot}$ (prompt
carbon ignition).
}
\label{fig:binary}
\end{figure}

In what binary systems ($P$ and $M_2$) does a uniformly rotating WD undergo
the delayed carbon ignition?  The result for the initial WD
masses of $0.9 M_{\odot}$ is shown in Figure \ref{fig:binary}.  Here
the binary systems starting from the ``painted'' region of the ($P -
M_2$) plane reach $1.38~M_{\odot} < M < 1.43~M_{\odot}$, while the
systems starting from the blank region encircled by the solid line
reach $M = 1.43~M_{\odot}$.  The occurrence frequency of the delayed
carbon ignition would roughly be one-third of the total frequency of
the carbon ignition.

For the values of $\tau_{\rm J}$ considered here, the WD spends enough time to
undergo SN~Ia explosion for the donor star to evolve to a
structure completely different from the one it had when acted as a
donor.  For the red-giant donor, its H-rich envelope would be lost as
a result of H-shell burning and mass loss so that it would become a He
WD in $\sim$ 10 Myr.  For the main sequence donor, it would also
evolve to become a low mass He WD in 1 Myr, a hot He WD in 10 Myr, and
a cold He WD in 1000 Myr \cite[]{2011ApJ...738L...1D}. So, the J-losses
should delay the explosion long enough for the former donor to be
undetectable. Therefore, this scenario provides a way to account for
the failure in detecting companions to SNe~Ia.

Such He white dwarf companions would be faint enough not to be seen
before or after the Type Ia supernova explosion. This new
single-degenerate scenario can explain in a unified manner why no
signatures of the companion star are seen in some Type Ia supernovae,
whereas some Type Ia supernovae indicate the presence of the companion
star.

\subsubsection{Differential Rotation and Super-Chandra SNe Ia} 
\index{Differential rotation} 

If the accretion leads to non-uniform, differentially rotating WDs, 
carbon ignition occurs at super-Chandrasekhar masses \cite[]{hachi12}. 
The WD mass can increase by accretion up to 2.3~(2.7)$~M_{\odot}$ from
the initial value of 1.1~(1.2)$~M_{\odot}$, being consistent with high
luminosity SNe~Ia such as SN~2003fg, SN~2006gz, SN~2007if, and
SN~2009dc \cite[]{Kamiya2012}.  Such very bright super-Chandrasekhar
mass SNe~Ia are suggested to born in a low metallicity environment.

\begin{figure}
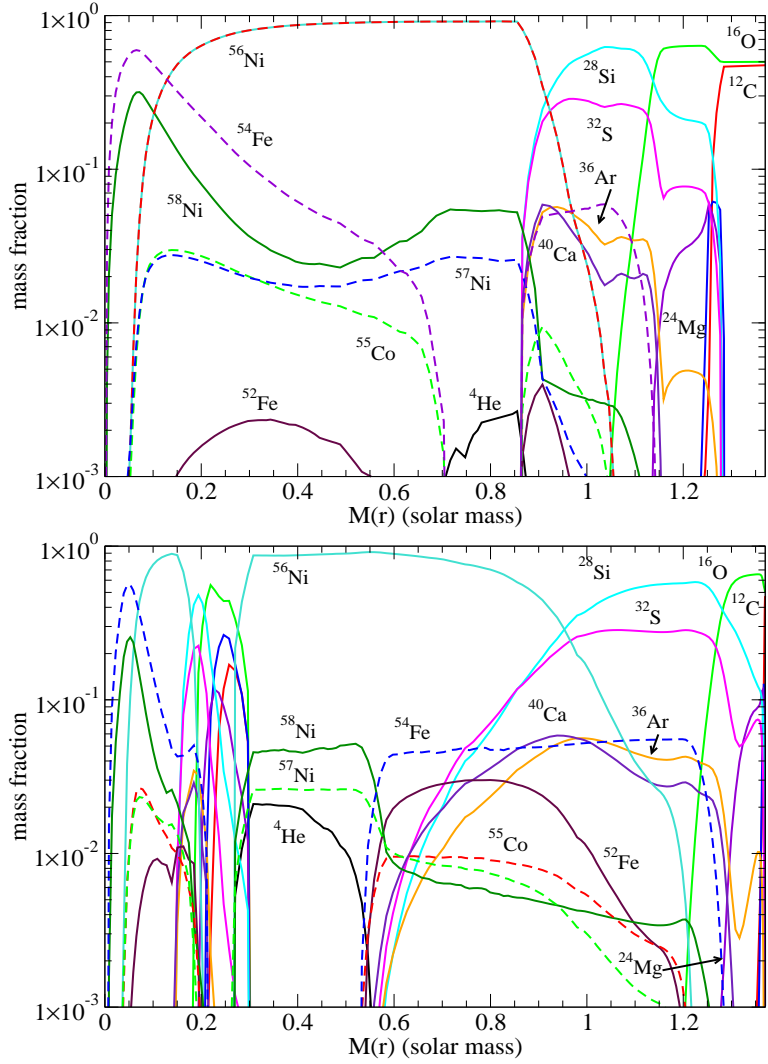

\centering
\includegraphics*[width=10cm,height=7cm]{fig16a.eps}
\includegraphics*[width=10cm,height=7cm]{fig16b.eps}
\caption{(upper panel) Mass fraction against $M_r$ for the W7 model
using the new nuclear reaction rates and electron
capture rates \cite[]{LN2017}. 
(lower panel) Same as the upper panel, but for the WDD2
model.  }
\label{fig:mr_plot}
\end{figure}

\begin{figure}
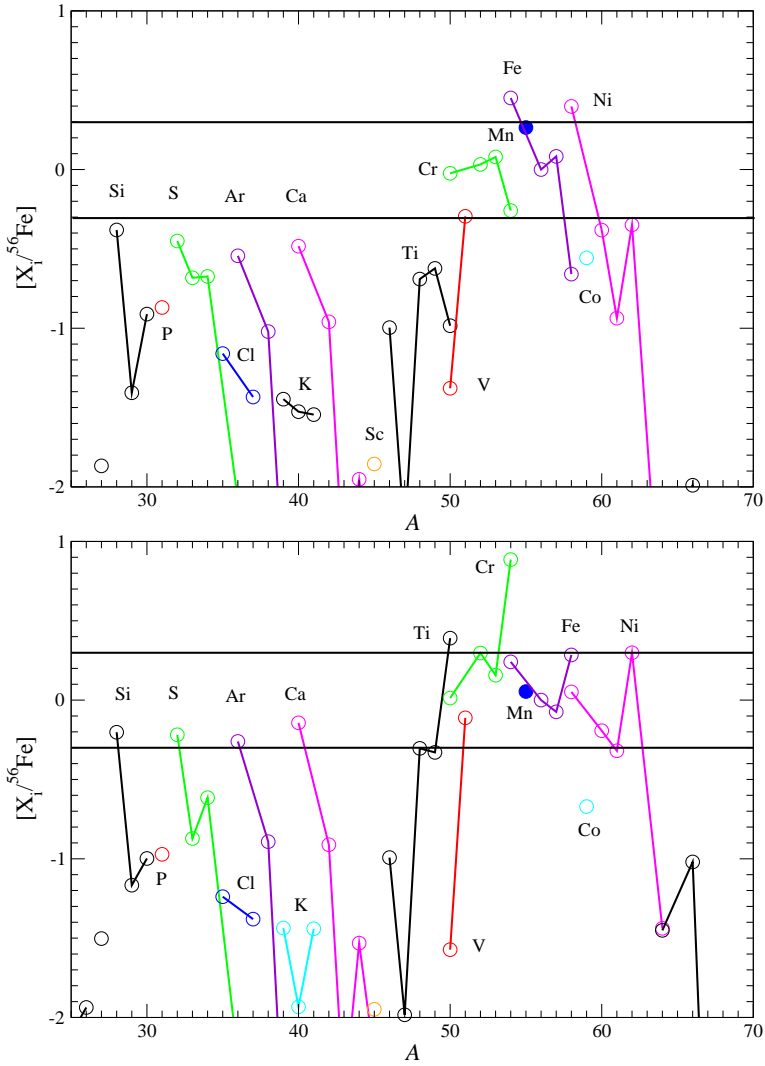

\centering
\includegraphics*[width=10cm,height=7cm]{fig16a_erratum.eps}
\includegraphics*[width=10cm,height=7cm]{fig16b_erratum.eps}
\caption{(upper panel) $[X_i/^{56}$Fe] for the W7 model
using the new nuclear reaction rates and electron
capture rates \cite[]{LN2017}. 
(lower panel) Same as the upper panel, but for the WDD2
model. The corresponding yielded mass for the stable and 
radioactive isotopes are also listed in Tables \ref{table:yield1}, \ref{table:yield2} and \ref{table:yield3}.
($Remark$: The two figures are replaced due to the updated table.)}
\label{fig:xiso_plot}
\end{figure}

\section{Nucleosynthesis Yields}

To distinguish the progenitor scenarios between SD and DD, or the the
Chandrasekhar mass and sub-Chandrasekhar mass models, comparisons of
nucleosynthesis features between these scenarios/models predictions
(e.g. \cite{Thielemann1986, nom13}).

To provide the basis to obtain nucleosynthesis constraints, we
recalculate nucleosynthesis yields of 1D explosion models with new
nuclear reaction rates and weak interaction rates 
(\cite{LN2017}; see also \cite{Mori2016}).

\subsection{Yields from Chandrasekhar Mass White Dwarfs}
\label{chandra}
\index{chandra}

We renew the nucleosynthesis yields of 1D models W7 and WDD2 \citep{Nomoto1984, Iwamoto1999}.
In these Chandrasekhar mass models, carbon burning ignited in the
central region is unstable to flash because of strong electron
degeneracy and release a large amount of nuclear energy explosively.
However, the central density is too high and thus the shock wave is
too weak to initiate a carbon detonation (because of
temperature-insensitive pressure of strongly degenerate electrons).

Then the explosive thermonuclear burning front propagates outward as a
convective deflagration wave (subsonic flame) \cite[]{Nomoto1976}.
Rayleigh-Taylor instabilities at the flame front cause the development
of turbulent eddies, which increase the flame surface area, enhancing
the net burning rate and accelerating the flame.  In the 1D convective
deflagration model W7 \cite[]{Nomoto1984}, the flame speed is
prescribed by time-dependent mixing-length theory with the mixing
length being 0.7 of the pressure scale height.  (In the central region,
the mixing length is assumed to be equal to the radial distance from
the center.)

In some cases the deflagration may undergo ``deflagration to
detonation transition (DDT)'' \cite[]{Khokhlov1991}.  In the 1D DDT
model WDD2 \cite[]{Iwamoto1999}, DDT is assumed to occur when the
density at the flame front decreases to 2 $\times 10^7$ g cm$^{-3}$.
Such a turbulent nature of the flame propagation has been studied in
multi-dimensional simulations \cite[e.g.,][]{Leung2015a, Leung2015b,Nomoto2017,LN2017}.

In Figure \ref{fig:mr_plot} we plot the abundance distributions (mass
fractions of main species) of W7 (upper panel) and WDD2 (lower panel).
In the central region of these models, the temperature behind the
deflagration wave exceeds \(\sim 5 \times 10^9\) K, so that the
reactions are rapid enough (compared with the expansion timescale) to
realize nuclear statistical equilibrium (NSE).  The central densities
of the WDs are so high ($\sim 3 \times 10^9$ g cm$^{-3}$) that
electron capture reduces the electron mole fraction, \(Y_e\), that is
the number of electrons per baryon. Then a significant amount of
neutron-rich species, such as \(^{58}\)Ni, \(^{57}\)Ni, \(^{56}\)Fe,
\(^{54}\)Fe, and \(^{55}\)Co (which decays to $^{55}$M), are
synthesized.

In W7, 0.65 $M_\odot$ $^{56}$Ni are produced.  The surrounding layers
at $M_r > 0.8 M_{\odot}$ gradually expand during the subsonic flame
propagation, so that the densities and temperatures get lower.  As a
result, explosive burning produces the intermediate mass elements
$^{28}$Si, $^{32}$S, $^{36}$Ar and $^{40}$Ca due to lower peak
temperatures than in the central region.  Beyond $M_r = 1.2 M_{\odot}$
explosive O-burning is slow so that $^{16}$O is dominant.  The
deflagration wave does not reach beyond $m(r) = 1.3 M_{\odot}$ so that
$^{12}$C and $^{16}$O remain unburned.

In WDD2, the deflagration speed is assumed to be slower than W7 and
undergoes the deflagration-detonation-transition (DDT).  Then the
chemical profile shows a two-layer structure.  In the innermost 0.3
$M_{\odot}$ where the materials are burnt by deflagration, dominant
products are similar to W7. After DDT, dominant nucleosythesis
products are somewhat neutron-rich Fe-peak species (e.g., $^{58}$Ni,
$^{57}$Ni, $^{56}$Ni) at $M_r <1 M_{\odot}$, intermediate mass
elements (e.g. $^{28}$Si and $^{32}$S) at $1 M_{\odot} < M_r < 1.3 M_{\odot}$,
and $^{16}$O in the outermost layers.  There is almost no
unburnt $^{12}$C, showing the detonation wave can sweep the whole
white dwarf before it quenches.  The total amount of $^{56}$Ni is 0.67
$M_\odot$ \cite[]{Iwamoto1999}.

Figures \ref{fig:xiso_plot} show $[X_i/^{56}$Fe] of the stable
isotopes from C to Zn for W7 (upper panel) and WDD2 (lower panel).
The detailed abundance ratios with respect to \(^{56}\)Fe depend on
the convective flame speed and the central densities, and also the
weak reaction rates.  In the old W7 and WDD2, electron capture rates
by \cite{Fuller1982} were used.  In Figure \ref{fig:xiso_plot}, the
most updated weak reaction rates are applied for electron capture
\cite{Langanke2001}.  The ratio of [$^{58}$Ni/$^{56}$Fe] is reduced
from $\sim 0.6$ in the old W7 to $\sim 0.3$ in the new W7, although
some neutron-rich species are still enhanced relative to $^{56}$Fe.

For W7, the metallicity dependent yields are calculated for the
initial mass fraction of $^{22}$Ne in the C+O white dwarfs
$X$($^{22}$Ne) = 0.025, 0.014, and 0.0025 as given in
\cite[]{yieldtable}.  The metallicity effects are not so large
in contrast to the sub-Chandrasekhar mass models (see next section)
because synthesis of neutron-rich species is mostly due to electron
capture in the NSE region rather than the initial metallicity
$X$($^{22}$Ne.

\begin{figure}[ht]
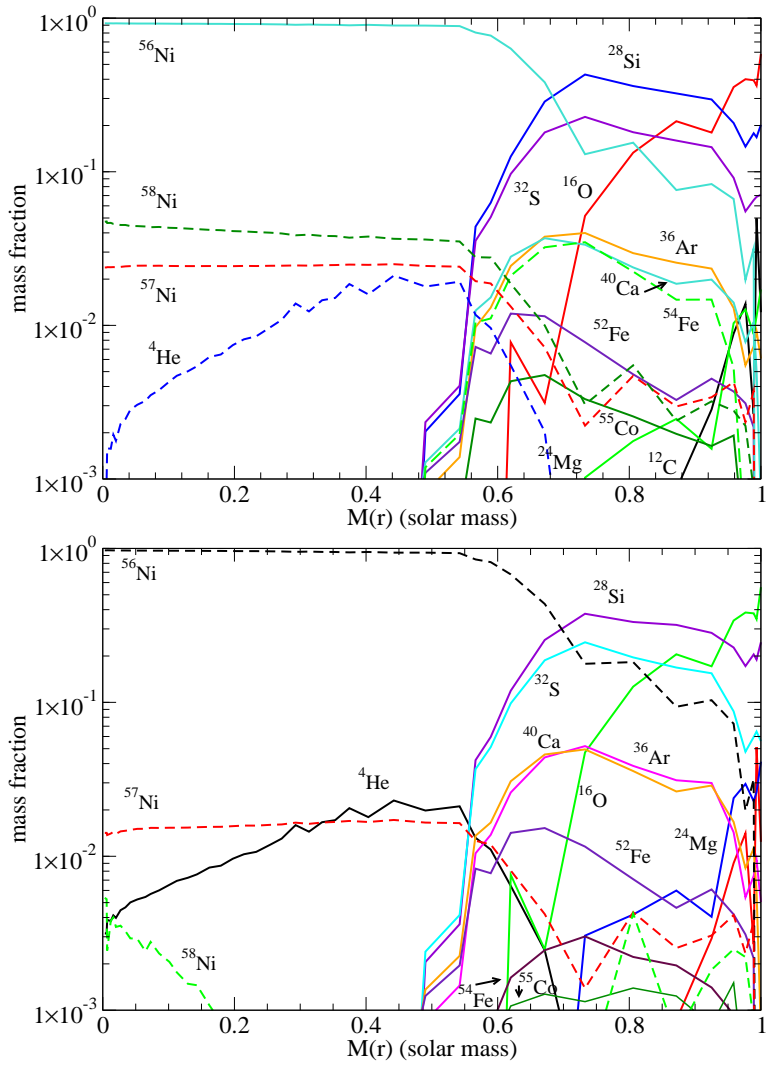

\begin{center}
\includegraphics*[width=10cm,height=7cm]{fig17a_new.eps}
\includegraphics*[width=10cm,height=7cm]{fig17b_new.eps}
\end{center}
\caption{
Abundance profiles
of the sub-Chandrasekhar mass model of 1.0 $M_\odot$ 
as a function of $M_r$ of the white dwarf.
}
\label{subChnuc}
\end{figure}

\begin{figure}
\begin{center}
\includegraphics*[scale=0.40]{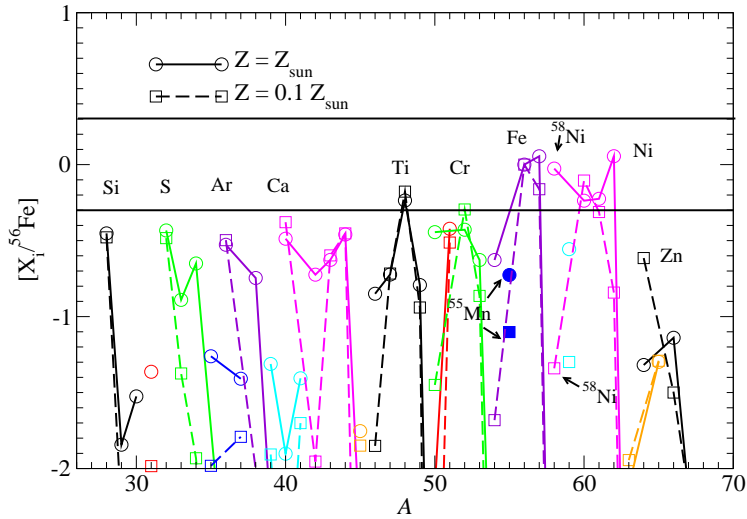}
\end{center}
\caption{
The ratios of integrated abundances of the sub-Chandrasekhar mass
model of 1.0 $M_\odot$ after the decay of unstable nuclei, normalized to
$^{56}$Fe, relative to solar abundances.  Models with $Z = Z_{\odot}$
(solid line) and $0.1 ~Z_{\odot}$ (dashed line) are used to contrast
the effects of metallicity.
($Remark$: The figure is replaced due to the updated table.)
}
\label{doubleDet}
\end{figure}

\subsection{Yields from sub-Chandrasekhar Mass White Dwarf Models}

There exist possible cases where an explosive carbon ignition occurs in
sub-Chandrasekhar mass C+O white dwarfs.  In the single degenerate
model, an off-center He detonation is induced by relatively slow
accretion of He as seen in Figure \ref{Heigmass}.  In some cases, a
resulting shock wave does not induce an immediate off-center
C-detonation but propagates toward the central region and increase its
strength because of decreasing area at the shock front.  Then the
shock wave becomes strong enough to ignite explosive carbon burning
which forms a detonation wave \cite[e.g.,][]{Livne1990, Woosley1994}.  In the
merging model, a merging of double white dwarfs ignites an off-center
carbon flash that produces a shock wave.  The shock wave propagates
through the center and induces C-detonation
\cite[e.g.,][]{Pakmor2011}.

\cite{Shigeyama1992} and \cite{Nomoto1994} calculated explosive
nucleosynthesis in such sub-Chandrasekhar mass models by artificially
inducing a C-detonation at center of the white dwarf.  We have
calculated similar sub-Chandrasekhar mass models by updating nuclear
reaction rates.  Figures \ref{subChnuc} show the abundance
distributions of the $M= 1.0 ~M_{\odot}$ white dwarf models for the
solar metallicity $Z = Z_{\odot}$ (upper) and the lower metallicity
$Z= 0.1 ~Z_{\odot}$ (lower).  For both models, $^{56}$Ni mass is
$\sim$ 0.6 $M_\odot$.

In the Fe-peak region, some neutron-rich species (\(^{58}\)Ni and
\(^{57}\)Ni) are synthesized because initial CNO elements are
converted to neutron-rich $^{22}$Ne during He-burning.  It is seen
that the amount of $^{58}$Ni in the low metallicity model is much
smaller than the solar metallicity model.

Figure \ref{doubleDet} shows integrated abundances of [X$_i/^{56}$Fe]
for both metallicity.  Because of the low neutron excess in the
central region, there is no overproduction of [X$_i/^{56}$Fe], in
particular, $^{54}$Fe, $^{58}$Ni, and $^{64}$Cr.  In fact, for the
solar metallicity model, [$^{58}$Ni/$^{56}$Fe] $\sim 0$.  However,
large underproduction is seen in [$^{55}$Mn/$^{56}$Fe] for both
metallicities and [$^{58}$Ni/$^{56}$Fe] for $Z= 0.1 ~Z_{\odot}$.  These
underproductions are important difference between the Chandrasekhar mass
and sub-Chandrasekhar mass models when compared with the observations.

For intermediate mass elements, there are almost no metallicity effects
on $^{28}$Si, $^{32}$S, $^{36}$Ar and $^{40}$Ca.  However,
underproduction is seen in slightly neutron-rich isotopes such as
$^{33}$S, $^{38}$Ar and $^{42}$Ca for $Z= 0.1 Z_{\odot}$.  These
species have a closer $Y_e$ to $^{22}$Ne.

In short, the yields of the sub-Chandrasekhar mass models with the
solar metallicity are similar to the Chandrasekhar mass model.  The
important difference, however, is the Mn yield, which is much smaller
in the sub-Chandrasekhar model than the Chandrasekhar mass models.
For the smaller metallicity, the differences are larger, so that the
comparison with observations of SNe and SNRs in small galaxies is
important.

\begin{figure}
\begin{center}
\includegraphics*[scale=0.40]{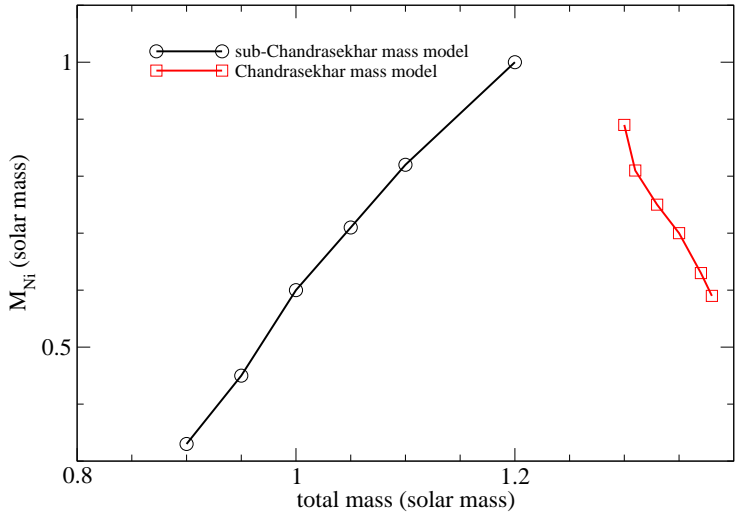}
\end{center}
\caption{
The $^{56}$Ni mass against white dwarf mass for the sub-Chandrasekhar
mass models and the near-Chandrasekhar mass models. Solar metallicity
is assumed for all models shown here.  For the sub-Chandrasekhar mass
models, the double detonation model, i.e., the He envelope with the
minimal mass required to trigger the carbon detonation is used.  For
Chandrasekhar mass models, the turbulent deflagration model with
deflagration-detonation transition is used \cite[]{LN2017}.
}
\label{mni_vs_m}
\end{figure}

\subsection{$^{56}$Ni Mass as a Function of White Dwarf Mass $M$}

It is interesting to know the maximum $M_{\rm Ni}$ which can be
produced in the explosion of WDs.  Figure \ref{mni_vs_m} shows the
$^{56}$Ni mass ($M_{\rm Ni}$) against the white dwarf mass $M$ for the
sub-Chandrasekhar mass models with solar metallicity.  The double
detonation model is used as the explosion mechanism for the
sub-Chandrasekhar mass white dwarfs, i.e., we consider carbon
detonation model triggered by spherical helium detonation.  Due to the
extra degree of freedom for the He envelope in the sub-Chandrasekhar
model, models with the minimal He envelope mass sufficient for
triggering the carbon detonation are used.

By symmetry, the spherical shock wave can propagate into the the
carbon-oxygen core without significant shock compression in the low
density region. As a result, the shock can reach deep in the core
where shock convergence can be significant to heat up the core and
trigger the carbon detonation.  

In the sub-Chandrasekhar mass model sequence, $M_{\rm Ni}$
sharply increases with $M$ for all models.  Models with $M \sim 1
M_{\odot}$ gives a typical $M_{\rm Ni}$ of 0.6 $M_{\odot}$.  This
corresponds to the bare CO mass of about 0.95 $M_{\odot}$.

There exist two factors that could affect the monotonic increase in
$M_{{\rm Ni}}$ with $M$.  One is the asphericity.  The aspherical
model always produces less $^{56}$Ni than the spherical because the
off-center carbon detonation allows more expansion before the
detonation wave sweeps across the whole white dwarf.

The other is the increasing central density with $M$.  At high central
density of the near-Chandrasekhar mass model, the shock wave is too
weak to form a detonation, so that a convective deflagration wave is
formed as discussed in \S \ref{chandra}.

Figure \ref{mni_vs_m} shows the $M_{\rm Ni}$ vs. $M$ relation for the
near-Chandrasekhar mass models.  Here the turbulent deflagration model
with deflagration-detonation transition (DDT) is used \cite[]{LN2017}.
There is a monotonic decrease with mass for the Chandrasekhar mass
sequence.  Since DDT occurs at sufficiently low density, a fraction of
the detonation that tends to produce larger $M_{\rm Ni}$ than
the deflagration is smaller for larger $M$.  (Even at higher central
densities, electron capture suppresses the production of $^{56}$Ni.)

As seen in Figure \ref{mni_vs_m}, the maximum $M_{\rm Ni}$ depends on
at which $M$, a deflagration rather than a detonation is formed in the
center.


\begin{table*}
\begin{center}
\caption{The yield table of the stable isotopes for 
the W7, WDD2 and the 1.0 $M_\odot$ sub-Chandrasekhar mass
models with $Z = Z_{\odot}$ and $Z = 0.1 ~Z_{\odot}$.
All short-lived isotopes are assumed to have decayed. 
All masses are in units of solar mass.
($Remark$: The table is updated and is different from the published version due to previous typos 
in the tables. 
)
}
\begin{tabular}{|c|c|c|c|c|c|}
 \hline
 Isotope & W7 & W7 & WDD2 & Sub-Chand. & Sub-Chand. \\ \hline
 Metallicity & $Z_{\odot}$ & 0.1 $Z_{\odot}$ & $Z_{\odot}$ & $Z_{\odot}$ & 0.1 $Z_{\odot}$ \\ \hline
 
$^{12}$C & $5.20 \times 10^{-2}$ & $5.44 \times 10^{-2}$ & $1.0 \times 10^{-2}$ & $1.15 \times 10^{-3}$ & $1.17 \times 10^{-3}$ \\
 $^{13}$C & $1.81 \times 10^{-11}$ & $1.37 \times 10^{-12}$ & $2.8 \times 10^{-7}$ & $3.2 \times 10^{-9}$ & $8.4 \times 10^{-11}$ \\
 $^{14}$N & $9.56 \times 10^{-10}$ & $8.11 \times 10^{-10}$ & $2.0 \times 10^{-7}$ & $1.81 \times 10^{-8}$ & $1.76 \times 10^{-9}$ \\
 $^{15}$N & $1.35 \times 10^{-10}$ & $1.41 \times 10^{-8}$ & $1.27 \times 10^{-8}$ & $2.13 \times 10^{-10}$ & $4.93 \times 10^{-9}$ \\
 $^{16}$O & $1.85 \times 10^{-1}$ & $1.35 \times 10^{-1}$ & $9.94 \times 10^{-2}$ & $6.64 \times 10^{-2}$ & $6.34 \times 10^{-2}$ \\
 $^{17}$O & $1.75 \times 10^{-10}$ & $1.27 \times 10^{-11}$ & $6.88 \times 10^{-8}$ & $1.8 \times 10^{-8}$ & $2.70 \times 10^{-10}$ \\
 $^{18}$O & $7.4 \times 10^{-12}$ & $7.45 \times 10^{-13}$ & $3.46 \times 10^{-9}$ & $9.43 \times 10^{-11}$ & $8.33 \times 10^{-12}$ \\
 $^{19}$F & $4.54 \times 10^{-12}$ & $3.89 \times 10^{-12}$ & $4.22 \times 10^{-10}$ & $2.39 \times 10^{-11}$ & $7.31 \times 10^{-12}$ \\
 $^{20}$Ne & $1.62 \times 10^{-3}$ & $1.64 \times 10^{-3}$ & $1.54 \times 10^{-2}$ & $1.15 \times 10^{-3}$ & $1.27 \times 10^{-3}$ \\
 $^{21}$Ne & $6.97 \times 10^{-8}$ & $3.16 \times 10^{-9}$ & $2.41 \times 10^{-6}$ & $1.57 \times 10^{-7}$ & $5.41 \times 10^{-9}$ \\
 $^{22}$Ne & $2.73 \times 10^{-3}$ & $2.73 \times 10^{-4}$ & $1.38 \times 10^{-5}$ & $9.25 \times 10^{-9}$ & $3.38 \times 10^{-9}$ \\
 $^{23}$Na & $6.61 \times 10^{-6}$ & $2.56 \times 10^{-6}$ & $1.47 \times 10^{-4}$ & $8.39 \times 10^{-6}$ & $3.86 \times 10^{-6}$ \\
 $^{24}$Mg & $4.26 \times 10^{-3}$ & $9.74 \times 10^{-3}$ & $1.3 \times 10^{-2}$ & $1.28 \times 10^{-3}$ & $3.10 \times 10^{-3}$ \\
 $^{25}$Mg & $1.81 \times 10^{-5}$ & $2.78 \times 10^{-6}$ & $2.98 \times 10^{-4}$ & $1.45 \times 10^{-5}$ & $8.53 \times 10^{-7}$ \\
 $^{26}$Mg & $2.35 \times 10^{-5}$ & $2.96 \times 10^{-6}$ & $5.2 \times 10^{-4}$ & $2.89 \times 10^{-5}$ & $3.45 \times 10^{-6}$ \\
 $^{26}$Al & $2.60 \times 10^{-10}$ & $5.4 \times 10^{-28}$ & $6.9 \times 10^{-9}$ & $2.60 \times 10^{-29}$ & $2.60 \times 10^{-29}$ \\
 $^{27}$Al & $4.50 \times 10^{-4}$ & $1.75 \times 10^{-4}$ & $1.2 \times 10^{-3}$ & $1.14 \times 10^{-4}$ & $4.76 \times 10^{-5}$ \\
 $^{28}$Si & $1.55 \times 10^{-1}$ & $1.43 \times 10^{-1}$ & $2.29 \times 10^{-1}$ & $1.25 \times 10^{-1}$ & $1.18 \times 10^{-1}$ \\
 $^{29}$Si & $7.73 \times 10^{-4}$ & $1.91 \times 10^{-4}$ & $1.31 \times 10^{-3}$ & $2.67 \times 10^{-4}$ & $9.34 \times 10^{-5}$ \\
 $^{30}$Si & $1.64 \times 10^{-3}$ & $7.10 \times 10^{-5}$ & $1.32 \times 10^{-3}$ & $3.79 \times 10^{-4}$ & $2.23 \times 10^{-5}$ \\
 $^{31}$P & $3.94 \times 10^{-4}$ & $5.76 \times 10^{-5}$ & $3.4 \times 10^{-4}$ & $1.53 \times 10^{-4}$ & $2.88 \times 10^{-5}$ \\
 $^{32}$S & $7.80 \times 10^{-2}$ & $8.29 \times 10^{-2}$ & $1.30 \times 10^{-1}$ & $6.56 \times 10^{-2}$ & $6.89 \times 10^{-2}$ \\
 $^{33}$S & $3.77 \times 10^{-4}$ & $1.16 \times 10^{-4}$ & $2.38 \times 10^{-4}$ & $1.89 \times 10^{-4}$ & $7.31 \times 10^{-5}$ \\
 $^{34}$S & $2.20 \times 10^{-3}$ & $1.17 \times 10^{-4}$ & $2.46 \times 10^{-3}$ & $1.91 \times 10^{-3}$ & $1.15 \times 10^{-4}$ \\
 $^{36}$S & $2.94 \times 10^{-7}$ & $2.25 \times 10^{-9}$ & $1.93 \times 10^{-7}$ & $6.35 \times 10^{-8}$ & $5.80 \times 10^{-10}$ \\
 $^{35}$Cl & $1.25 \times 10^{-4}$ & $1.73 \times 10^{-5}$ & $1.2 \times 10^{-4}$ & $1.4 \times 10^{-4}$ & $1.80 \times 10^{-5}$ \\
 $^{37}$Cl & $2.28 \times 10^{-5}$ & $8.36 \times 10^{-6}$ & $2.53 \times 10^{-5}$ & $2.51 \times 10^{-5}$ & $9.59 \times 10^{-6}$ \\
 $^{36}$Ar & $1.33 \times 10^{-2}$ & $1.77 \times 10^{-2}$ & $2.50 \times 10^{-2}$ & $1.15 \times 10^{-2}$ & $1.41 \times 10^{-2}$ \\
 $^{38}$Ar & $8.75 \times 10^{-4}$ & $5.15 \times 10^{-5}$ & $1.15 \times 10^{-3}$ & $1.34 \times 10^{-3}$ & $8.62 \times 10^{-5}$ \\
 $^{40}$Ar & $7.39 \times 10^{-9}$ & $1.55 \times 10^{-11}$ & $3.18 \times 10^{-9}$ & $1.71 \times 10^{-9}$ & $4.26 \times 10^{-12}$ \\
 $^{39}$K & $6.57 \times 10^{-5}$ & $1.18 \times 10^{-5}$ & $6.59 \times 10^{-5}$ & $9.25 \times 10^{-5}$ & $2.17 \times 10^{-5}$ \\
 $^{40}$K & $8.27 \times 10^{-8}$ & $2.0 \times 10^{-9}$ & $3.17 \times 10^{-8}$ & $3.84 \times 10^{-8}$ & $1.58 \times 10^{-9}$ \\
 $^{41}$K & $4.8 \times 10^{-6}$ & $1.45 \times 10^{-6}$ & $5.7 \times 10^{-6}$ & $5.64 \times 10^{-6}$ & $2.71 \times 10^{-6}$ \\
 $^{40}$Ca & $1.15 \times 10^{-2}$ & $1.74 \times 10^{-2}$ & $2.47 \times 10^{-2}$ & $1.4 \times 10^{-2}$ & $1.40 \times 10^{-2}$ \\
 $^{42}$Ca & $2.66 \times 10^{-5}$ & $1.49 \times 10^{-6}$ & $2.92 \times 10^{-5}$ & $4.24 \times 10^{-5}$ & $2.58 \times 10^{-6}$ \\
 $^{43}$Ca & $1.49 \times 10^{-7}$ & $4.70 \times 10^{-8}$ & $1.65 \times 10^{-7}$ & $1.38 \times 10^{-5}$ & $1.36 \times 10^{-5}$ \\
 $^{44}$Ca & $9.17 \times 10^{-6}$ & $1.42 \times 10^{-5}$ & $2.36 \times 10^{-5}$ & $2.63 \times 10^{-4}$ & $2.76 \times 10^{-4}$ \\
 $^{46}$Ca & $3.26 \times 10^{-10}$ & $4.35 \times 10^{-11}$ & $1.40 \times 10^{-9}$ & $4.90 \times 10^{-11}$ & $8.14 \times 10^{-16}$ \\
 $^{48}$Ca & $2.19 \times 10^{-12}$ & $2.9 \times 10^{-12}$ & $1.36 \times 10^{-9}$ & $9.28 \times 10^{-16}$ & $1.61 \times 10^{-22}$ \\
 $^{45}$Sc & $2.81 \times 10^{-7}$ & $9.74 \times 10^{-8}$ & $2.22 \times 10^{-7}$ & $3.77 \times 10^{-7}$ & $2.72 \times 10^{-7}$ \\
 $^{46}$Ti & $1.27 \times 10^{-5}$ & $8.69 \times 10^{-7}$ & $1.26 \times 10^{-5}$ & $1.85 \times 10^{-5}$ & $1.70 \times 10^{-6}$ \\
 $^{47}$Ti & $5.74 \times 10^{-7}$ & $1.39 \times 10^{-7}$ & $1.21 \times 10^{-6}$ & $2.27 \times 10^{-5}$ & $2.18 \times 10^{-5}$ \\
 $^{48}$Ti & $2.51 \times 10^{-4}$ & $3.66 \times 10^{-4}$ & $5.99 \times 10^{-4}$ & $7.8 \times 10^{-4}$ & $7.82 \times 10^{-4}$ \\
 $^{49}$Ti & $2.23 \times 10^{-5}$ & $1.53 \times 10^{-5}$ & $4.29 \times 10^{-5}$ & $1.47 \times 10^{-5}$ & $1.2 \times 10^{-5}$ \\
 $^{50}$Ti & $9.60 \times 10^{-6}$ & $9.33 \times 10^{-6}$ & $2.22 \times 10^{-4}$ & $6.7 \times 10^{-10}$ & $1.58 \times 10^{-13}$ \\
 $^{50}$V & $1.84 \times 10^{-8}$ & $6.94 \times 10^{-9}$ & $1.15 \times 10^{-8}$ & $3.68 \times 10^{-9}$ & $2.22 \times 10^{-11}$ \\
 $^{51}$V & $9.47 \times 10^{-5}$ & $5.84 \times 10^{-5}$ & $1.41 \times 10^{-4}$ & $7.62 \times 10^{-5}$ & $5.47 \times 10^{-5}$ \\

\end{tabular}
\label{table:yield1}
\end{center}
\end{table*}

\begin{table*}
\begin{center}
\caption{(cont'd) The yield table of the stable isotopes for 
the W7, WDD2 and the 1.0 $M_\odot$ sub-Chandrasekhar mass
models with $Z = Z_{\odot}$ and $Z = 0.1 ~Z_{\odot}$.
All masses are in units of solar mass.
($Remark$: The table is updated and is different from the published version due to previous typos 
in the tables. 
)
}
\begin{tabular}{|c|c|c|c|c|c|}
 \hline
 Isotope & W7 & W7 & WDD2 & Sub-Chand. & Sub-Chand. \\ \hline
 Metallicity & $Z_{\odot}$ & 0.1 $Z_{\odot}$ & $Z_{\odot}$ & $Z_{\odot}$ & 0.1 $Z_{\odot}$ \\ \hline
 $^{50}$Cr & $3.74 \times 10^{-4}$ & $1.54 \times 10^{-4}$ & $3.99 \times 10^{-4}$ & $1.41 \times 10^{-4}$ & $1.34 \times 10^{-5}$ \\
 $^{52}$Cr & $8.59 \times 10^{-3}$ & $1.3 \times 10^{-2}$ & $1.54 \times 10^{-2}$ & $2.94 \times 10^{-3}$ & $3.86 \times 10^{-3}$ \\
 $^{53}$Cr & $1.11 \times 10^{-3}$ & $8.73 \times 10^{-4}$ & $1.30 \times 10^{-3}$ & $2.14 \times 10^{-4}$ & $1.21 \times 10^{-4}$ \\
 $^{54}$Cr & $1.30 \times 10^{-4}$ & $1.26 \times 10^{-4}$ & $1.77 \times 10^{-3}$ & $4.40 \times 10^{-8}$ & $1.18 \times 10^{-9}$ \\
 $^{55}$Mn & $1.36 \times 10^{-2}$ & $1.3 \times 10^{-2}$ & $8.21 \times 10^{-3}$ & $1.32 \times 10^{-3}$ & $5.59 \times 10^{-4}$ \\
 $^{54}$Fe & $1.15 \times 10^{-1}$ & $8.82 \times 10^{-2}$ & $6.98 \times 10^{-2}$ & $8.69 \times 10^{-3}$ & $8.14 \times 10^{-4}$ \\
 $^{56}$Fe & $6.68 \times 10^{-1}$ & $7.24 \times 10^{-1}$ & $6.54 \times 10^{-1}$ & $6.0 \times 10^{-1}$ & $6.38 \times 10^{-1}$ \\
 $^{57}$Fe & $1.96 \times 10^{-2}$ & $1.51 \times 10^{-2}$ & $1.34 \times 10^{-2}$ & $1.60 \times 10^{-2}$ & $1.6 \times 10^{-2}$ \\
 $^{58}$Fe & $5.46 \times 10^{-4}$ & $5.32 \times 10^{-4}$ & $4.70 \times 10^{-3}$ & $1.36 \times 10^{-8}$ & $1.86 \times 10^{-9}$ \\
 $^{60}$Fe & $7.33 \times 10^{-10}$ & $6.97 \times 10^{-10}$ & $4.10 \times 10^{-8}$ & $3.4 \times 10^{-18}$ & $1.81 \times 10^{-21}$ \\
 $^{59}$Co & $5.20 \times 10^{-4}$ & $4.29 \times 10^{-4}$ & $3.92 \times 10^{-4}$ & $5.27 \times 10^{-4}$ & $9.1 \times 10^{-5}$ \\
 $^{58}$Ni & $6.80 \times 10^{-2}$ & $4.71 \times 10^{-2}$ & $3.0 \times 10^{-2}$ & $2.50 \times 10^{-2}$ & $1.18 \times 10^{-3}$ \\
 $^{60}$Ni & $4.51 \times 10^{-3}$ & $5.39 \times 10^{-3}$ & $6.82 \times 10^{-3}$ & $6.13 \times 10^{-3}$ & $8.14 \times 10^{-3}$ \\
 $^{61}$Ni & $5.81 \times 10^{-5}$ & $5.77 \times 10^{-5}$ & $2.35 \times 10^{-4}$ & $2.78 \times 10^{-4}$ & $2.33 \times 10^{-4}$ \\
 $^{62}$Ni & $7.3 \times 10^{-4}$ & $3.83 \times 10^{-4}$ & $3.5 \times 10^{-3}$ & $1.72 \times 10^{-3}$ & $2.15 \times 10^{-4}$ \\
 $^{64}$Ni & $4.17 \times 10^{-7}$ & $4.6 \times 10^{-7}$ & $1.70 \times 10^{-5}$ & $4.59 \times 10^{-14}$ & $3.79 \times 10^{-14}$ \\
 $^{63}$Cu & $5.15 \times 10^{-7}$ & $3.57 \times 10^{-7}$ & $1.69 \times 10^{-6}$ & $2.50 \times 10^{-6}$ & $3.69 \times 10^{-6}$ \\
 $^{65}$Cu & $1.96 \times 10^{-7}$ & $1.96 \times 10^{-7}$ & $1.3 \times 10^{-6}$ & $7.61 \times 10^{-6}$ & $7.68 \times 10^{-6}$ \\
 $^{64}$Zn & $1.34 \times 10^{-6}$ & $4.20 \times 10^{-6}$ & $1.96 \times 10^{-5}$ & $2.70 \times 10^{-5}$ & $1.31 \times 10^{-4}$ \\
 $^{66}$Zn & $3.42 \times 10^{-6}$ & $1.14 \times 10^{-6}$ & $3.12 \times 10^{-5}$ & $2.41 \times 10^{-5}$ & $1.1 \times 10^{-5}$ \\
 $^{67}$Zn & $2.28 \times 10^{-9}$ & $4.24 \times 10^{-10}$ & $1.90 \times 10^{-8}$ & $3.41 \times 10^{-7}$ & $3.33 \times 10^{-7}$ \\
 $^{68}$Zn & $1.29 \times 10^{-9}$ & $2.95 \times 10^{-9}$ & $1.61 \times 10^{-8}$ & $3.71 \times 10^{-7}$ & $6.8 \times 10^{-7}$ \\
 $^{70}$Zn & $2.93 \times 10^{-14}$ & $2.85 \times 10^{-14}$ & $1.29 \times 10^{-11}$ & $1.97 \times 10^{-18}$ & $1.48 \times 10^{-18}$ \\
 \hline

\end{tabular}
\label{table:yield2}
\end{center}
\end{table*}

\begin{table*}
\begin{center}
\caption{The yield table of the radioactive isotopes for 
the W7, WDD2 and the 1.0 $M_\odot$ sub-Chandrasekhar mass
models with $Z = Z_{\odot}$ and $Z = 0.1 ~Z_{\odot}$.
All masses are in units of solar mass.
($Remark$: The table is updated and is different from the published version due to previous typos 
in the tables.)}
\begin{tabular}{|c|c|c|c|c|c|}
 \hline
 Isotope & W7 & W7 & WDD2 & Sub-Chand. & Sub-Chand. \\ \hline
 Metallicity & $Z_{\odot}$ & 0.1 $Z_{\odot}$ & $Z_{\odot}$ & $Z_{\odot}$ & 0.1 $Z_{\odot}$ \\ \hline
 $^{22}$Na & $4.89 \times 10^{-9}$ & $2.89 \times 10^{-9}$ & $6.20 \times 10^{-8}$ & $3.74 \times 10^{-9}$ & $3.22 \times 10^{-9}$ \\
 $^{26}$Al & $2.23 \times 10^{-6}$ & $1.66 \times 10^{-6}$ & $2.84 \times 10^{-5}$ & $1.70 \times 10^{-6}$ & $1.74 \times 10^{-6}$ \\
 $^{39}$Ar & $1.71 \times 10^{-8}$ & $2.20 \times 10^{-10}$ & $6.11 \times 10^{-9}$ & $7.27 \times 10^{-9}$ & $1.11 \times 10^{-10}$ \\
 $^{40}$K & $8.32 \times 10^{-8}$ & $2.1 \times 10^{-9}$ & $3.19 \times 10^{-8}$ & $3.86 \times 10^{-8}$ & $1.59 \times 10^{-9}$ \\
 $^{41}$Ca & $4.5 \times 10^{-6}$ & $1.47 \times 10^{-6}$ & $4.99 \times 10^{-6}$ & $5.66 \times 10^{-6}$ & $2.38 \times 10^{-6}$ \\
 $^{44}$Ti & $8.48 \times 10^{-6}$ & $1.33 \times 10^{-5}$ & $2.21 \times 10^{-5}$ & $2.64 \times 10^{-4}$ & $2.68 \times 10^{-4}$ \\
 $^{48}$V & $4.24 \times 10^{-8}$ & $1.21 \times 10^{-8}$ & $6.88 \times 10^{-8}$ & $1.96 \times 10^{-7}$ & $1.47 \times 10^{-7}$ \\
 $^{49}$V & $2.21 \times 10^{-7}$ & $3.59 \times 10^{-8}$ & $1.25 \times 10^{-7}$ & $1.54 \times 10^{-7}$ & $1.18 \times 10^{-8}$ \\
 $^{53}$Mn & $1.97 \times 10^{-4}$ & $1.75 \times 10^{-4}$ & $1.44 \times 10^{-4}$ & $9.5 \times 10^{-6}$ & $4.27 \times 10^{-7}$ \\
 $^{60}$Fe & $1.2 \times 10^{-8}$ & $9.99 \times 10^{-9}$ & $5.73 \times 10^{-7}$ & $4.59 \times 10^{-17}$ & $2.90 \times 10^{-20}$ \\
 $^{56}$Co & $1.8 \times 10^{-4}$ & $8.97 \times 10^{-5}$ & $5.60 \times 10^{-5}$ & $8.93 \times 10^{-6}$ & $5.89 \times 10^{-6}$ \\
 $^{57}$Co & $7.65 \times 10^{-4}$ & $7.32 \times 10^{-4}$ & $3.48 \times 10^{-4}$ & $5.62 \times 10^{-6}$ & $1.23 \times 10^{-6}$ \\
 $^{60}$Co & $7.72 \times 10^{-8}$ & $7.49 \times 10^{-8}$ & $3.52 \times 10^{-7}$ & $1.17 \times 10^{-13}$ & $7.99 \times 10^{-15}$ \\
 $^{56}$Ni & $6.45 \times 10^{-1}$ & $7.2 \times 10^{-1}$ & $6.32 \times 10^{-1}$ & $6.0 \times 10^{-1}$ & $6.38 \times 10^{-1}$ \\
 $^{57}$Ni & $1.88 \times 10^{-2}$ & $1.42 \times 10^{-2}$ & $1.28 \times 10^{-2}$ & $1.60 \times 10^{-2}$ & $1.6 \times 10^{-2}$ \\
 $^{59}$Ni & $3.3 \times 10^{-4}$ & $2.86 \times 10^{-4}$ & $1.1 \times 10^{-4}$ & $2.89 \times 10^{-6}$ & $1.61 \times 10^{-6}$ \\
 $^{63}$Ni & $8.47 \times 10^{-8}$ & $8.26 \times 10^{-8}$ & $8.17 \times 10^{-7}$ & $2.12 \times 10^{-15}$ & $9.34 \times 10^{-16}$ \\
 \hline

\end{tabular}
\label{table:yield3}
\end{center}
\end{table*}

\section{Summary}

In this review, we review how the single degenerate models for Type Ia
supernovae (SNe Ia) work. In this binary system, the white dwarf (WD)
accretes either hydrogen-rich matter or helium and undergoes hydrogen
and helium shell-burning from its non-degenerate companion star.  We
summarize how the stability and non-linear behavior of such
shell-burning depend on the accretion rate and the WD mass and how the
WD blows strong wind.

We identify the following evolutionary routes for the accreting WD to
trigger a thermonuclear explosion.  Typically, the accretion rate is
quite high in the early stage and gradually decreases as a result of
mass transfer.  With decreasing rate, the WD evolves as follows:

\noindent
(1) At a rapid accretion phase, the WD increase its mass by stable H
burning and blows a strong wind to keep its moderate radius.  The wind
is strong enough to strip a part of the companion star's envelope to
control the accretion rate and forms circumstellar matter (CSM).  If
the WD explodes within CSM, it is observed as an "SN Ia-CSM".
(X-rays emitted by the WD are absorbed by CSM.)

\noindent
(2) If the WD continues to accrete at a lower rate, the wind stops and
an SN Ia is triggered under steady-stable H shell-burning, which is
observed as a super-soft X-ray source: "SN Ia-SXSS".

\noindent
(3) If the accretion continues at a still lower rate, H shell-burning
becomes unstable and many flashes recur.  The WD undergoes recurrent
nova (RN) whose mass ejection is smaller than the accreted matter.
Then the WD evolves to an "SN Ia-RN".

\noindent
(4) If the companion is a He star (or a He WD), the accretion of He
can trigger He and C double detonations at the sub-Chandrasekhar mass
or the WD grows to the Chandrasekhar mass under a He-wind: "SN Ia-He
CSM".

\noindent
(5) If the accreting WD rotates quite rapidly, the WD mass can exceed
the Chandrasekhar mass of the spherical WD, which delays the trigger
of an SN Ia.  After angular momentum is lost from the WD, the
(super-Chandra) WD contracts to become a delayed SN Ia.  The companion
star has become a He WD and CSM has disappeared: ``SN Ia-He WD''.

\noindent
Finally, we update nucleosynthesis yields of the carbon deflagration
model W7, delayed detonation model WDD2, and the sub-Chandrasekhar
mass model to provide some constraints on the yields (such as Mn) from
the comparison with the observations.  We note the important metallicity
effects on $^{58}$Ni and $^{55}$Mn.

\begin{acknowledgements}
This work has been supported by by the World Premier International
Research Center Initiative (WPI Initiative), MEXT, Japan, and JSPS
KAKENHI Grant Numbers JP26400222, JP16H02168, JP17K05382.
\end{acknowledgements}

\bibliographystyle{aps-nameyear}

\end{document}